\documentclass[12pt, preprint]{aastex}

\shorttitle{Atomic Diagnostics of X-ray Irradiated Protoplanetary Disks}
\shortauthors{Meijerink et al.}

\begin{document}

\title{Atomic Diagnostics of X-ray Irradiated Protoplanetary Disks}

\author{R. Meijerink and A. E. Glassgold}
\affil{Astronomy Department, University of California,
    Berkeley, CA 94720}
\email{rowin@astro.berkeley.edu, glassgol@astro.berkeley.edu}

\and

\author{J. R. Najita}
\affil{National Optical Astronomy Observatory, Tucson, AZ 85719}
\email{najita@noao.edu}

\begin{abstract}

We study atomic line diagnostics of the inner regions of
protoplanetary disks with our model of X-ray irradiated disk
atmospheres which was previously used to predict observable levels of
the Ne\,II and Ne\,III fine-structure transitions at 12.81 and
15.55\,$\mu$m. We extend the X-ray ionization theory to sulfur and
calculate the fraction of sulfur in S, S$^+$, S$^{2+}$ and sulfur
molecules. For the D'Alessio generic T Tauri star disk, we find that
the S\,I fine-structure line at 25.55\,$\mu$m is below the detection
level of the {\it Spitzer} Infrared Spectrometer (IRS), in large part
due to X-ray ionization of atomic S at the top of the atmosphere and
to its incorporation into molecules close to the mid-plane. We predict
that observable fluxes of the S\,II 6718/6732\,\AA\ forbidden
transitions are produced in the upper atmosphere at somewhat shallower
depths and smaller radii than the neon fine-structure lines. This and
other forbidden line transitions, such as the O\,I 6300/6363\,\AA\ and
the C\,I 9826/9852\,\AA\ lines, serve as complementary diagnostics of
X-ray irradiated disk atmospheres. We have also analyzed the potential
role of the low-excitation fine-structure lines of C\,I, C\,II, and
O\,I, which should be observable by SOFIA and {\it Herschel}.

\end{abstract}

\keywords{accretion, accretion disks -- infrared: stars -- 
planetary systems: protoplanetary disks -- stars: formation --
stars: pre-man-sequence -- X-rays: stars}

\section{Introduction}

There is growing evidence for the existence of warm atmospheres in the
inner disks of low and intermediate mass young stellar objects
(YSOs). They first became apparent from observations of the CO
overtone bandhead emission near 2.3 $\mu$m in Herbig Ae stars and T
Tauri stars (e.g., Carr \citeyear{Carr1989}; Najita et al.
\citeyear{Najita2000}). More recently, the emission from the
fundamental CO ro-vibrational band near 4.6 $\mu$m has been observed
in many T Tauri stars \citep{Najita2003}, as has the UV fluorescence
of molecular hydrogen (e.g., Herczeg et
al. \citeyear{Herczeg2002}). These observations indicate that
temperatures in the range 1000-2000\,K occur over a considerable
vertical column density of hydrogen $\sim 10^{21}$ cm$^{-2}$ in the
inner disk region. This and other spectroscopic evidence for warm
gaseous disk atmospheres were reviewed by \citet{Najita2007}, where
extensive references are given to earlier work.

\citet{Calvet1991} pioneered the now widely accepted view that disk
atmospheres are heated by stellar radiation. In previous publications
(Glassgold, Najita \& Igea \citeyear{Glassgold2004}, henceforth GNI04)
we argued that stellar X-rays play an important role in heating the
upper gas layers of circumstellar disks on the following basis. First,
significant X-ray emission appears to be a universal aspect of
low-mass YSOs; second, X-rays couple strongly to the gas; and third,
hard X-rays with energies $\gtrsim 1 $\,keV can penetrate deeply into
disk atmospheres.

One special aspect of X-ray irradiation is that temperatures as large
as 4000-5000\,K are achieved high in the atmosphere. We recently
pointed out that this hot layer gives rise to significant flux levels
of the fine-structure lines of neon at 12.81 $\mu$m and 15.55 $\mu$m
(Glassgold et al. \citeyear{Glassgold2007}, henceforth GNI07).

The Ne\,II 12.81 $\mu$m line has now been detected in many YSOs by the
{\it Spitzer} Infrared Spectrometer (IRS) in nearby star-forming
regions \citep{Pascucci2007, Lahuis2007, Espaillat2007} and by
MICHELLE at Gemini North \citep{Herczeg2007}. The measured fluxes are
in good agreement with the predictions in GNI07. These observations
support the hypothesis that disk atmospheres are heated and ionized by
photons energetic enough to ionize Ne (which has first and second
ionization potentials of 21.56 and 41.0 eV), most likely X-rays. A key
motivation in the work of GNI07 was the search for clear diagnostics
of the effects of X-ray irradiation.  The Ne lines seemed particularly
appropriate because hard X-rays generate high ionization states of
neon via the Auger effect and because neon chemistry is especially
simple. Gorti \& Hollenbach (2006, private communication) have
suggested that EUV radiation may contribute to the neon fluxes,
although the emission characteristics of YSOs in this wavelength band
are poorly known compared to X-rays.

In this paper we extend the considerations of GNI07 and seek other
fine-structure and forbidden transitions that have a potential to
complement the neon lines. We have been guided in part by the
intrinsic strength of a line, which we can take as the thermalized
emissivity in a transition from an upper to a lower level
$u \rightarrow l$,
\begin{equation}
j(u,l) = P(u)\ xn_{\rm H}\ A(u,l)\ E(u,l).
\label{thermalemission}
\end{equation}
Here $xn_{\rm H}$ is the density of the emitting species, $P(u)$ is
the population of the upper level, and $A(u,l)$ and $E(u,l)$ are the
Einstein $A$-value and energy of the transition. The last two factors
in equation~(\ref{thermalemission}) emphasize high frequency
transitions, but the other factors are crucial since they require a
theory for the ionization of the carrier species (the $x$ factor) and
the excitation of the level (the factor $P(u)$), since they are sensitive 
to the physical conditions. We have chosen to focus here on the lines of
oxygen, sulfur and, to a lesser extent, carbon. Some spectral lines of
particular interest are the 25.25\,$\mu$m fine-structure transition of
S\,I, which has been searched for, mostly without success, by {\it Spitzer};
the fine-structure lines of O\,I, which should be detectable with 
SOFIA and the PACS imaging spectrometer on {\it Herschel}; and various 
optical-infrared forbidden transitions accessible from the ground. 

The atomic carriers of these lines appear in the many disk chemical
models developed over the last 10-15 years (reviewed by Bergin et
al.~\citeyear{Bergin2007}). They are closely related to the
photon-dominated models (PDR) developed for the interstellar
medium. In this case stellar and interstellar radiation generate
warm atomic regions in the upper and lower layers of the disks until
the UV radiation is absorbed and molecule formation becomes
efficient. Jonkheid and Kamp and their collaborators
\citep{Jonkheid2004, Jonkheid2006, Jonkheid2007, Kamp2003} have gone
further and calculated the emissivity and lineshape of the
fine-structure lines of C\,I, C\,II, and O\,I.

The 63 $\mu$m fine-structure line of O\,I was detected in YSOs with
the KAO (e.g., Cohen et al.~\citeyear{Cohen1988}; Ceccarelli et
al.~\citeyear{Ceccarelli1997}) and with ISO (e.g., Nisini et
al.~\citeyear{Nisini1999}; Spinoglio et al.~\citeyear{Spinoglio2000};
Creech-Eakman et al.~\citeyear{Creech2002}; Liseau et
al.~\citeyear{Liseau2006}), but the interpretation of the results is
hampered by the low spatial and spectral resolution of the
observations so that it is unclear what fraction of the emission
arises from the disk. Disks have also been considered as a possible
source of forbidden line emission by \citet{Hartigan1995},
\citet{Kwan1997} and \citet{Stoerzer2000}. The forbidden lines have
been well studied in YSOs (see, e.g., the review by Ray et
al.~\citeyear{Ray2007}), and some of the observed
``low-velocity-components'' of the emission may originate from the
disks in these systems (e.g., Kwan \& Tademaru \citeyear{Kwan1988},
\citeyear{Kwan1995}). We will return to the discussion of the
observations of the fine-structure and forbidden lines in Section~5.

The calculations reported in this paper are exploratory in nature, as
were those in GNI07. They examine the ability of X-rays, a well
characterized property of TTS, to produce atomic line emission from T
Tauri disks. The model employs the continuous disk density
distribution developed by D'Alessio et al.~(1999) for a generic T
Tauri star, and they feature stellar X-ray heating and ionization in
the context of a simplified thermal-chemical model, which will be
discussed in more detail in Section 2. We do not attempt to fit the
observations for any specific system, but instead try to derive
conclusions that might be applicable to any protoplanetary disk.

The plan of the rest of this paper is as follows. In the next section,
we review and update the thermal-chemical model of GNI04. In Section
3, we summarize the essentials of the basic ionization theory and
extend it to include the X-ray ionization of sulfur. In Section 4, we
outline the excitation and flux calculations for the lines of
interest, and then discuss the relevant observations in Section 5. The
last section summarizes the main results of this paper.

\section{Review Of The Model}

We use the thermal-chemical model of GNI04 with minor corrections and
updates. The important processes in determining the thermal balance
are X-ray heating, gas-grain heating and cooling, mechanical heating
(e.g., from disk accretion or the wind-disk interaction), and line
cooling (Ly$\alpha$, recombination lines, O\,I fine-structure and
forbidden lines, and CO rotational and ro-vibrational lines). The
chemical network contains about 125 reactions and 25 species. Errors
in the specification of these processes have been corrected, and the
entire reaction base has been re-evaluated and updated. None of the
changes affects the conclusions of GNI04 and GNI07 in any major way.
GNI04 provide a detailed description of the physical processes, and
the chemical reaction base is available from the authors on request or
can be downloaded at
http://astro.berkeley.edu/$\sim$rowin/MGN$\_$rates.pdf.

Following GNI04 and GNI07, we calculate the chemical and thermal
structure of the gas in a protoplanetary disk for the smooth density
distribution of the generic T-Tauri disk model of \citet{dAlessio1999}
using an updated version of the code developed by
GNI04. \citet{dAlessio1999} did not distinguish between the thermal
properties of the dust and the gas, and their density structure is in
hydrostatic equilibrium. However, at high altitudes and low vertical
column densities, the temperature of the gas is much higher than the
dust (GNI04), and this implies that our model is no longer in
hydrostatic equilibrium.  Were we to treat the gas and dust as two
separate fluids and require hydrostatic equilibrium, the results would
differ somewhat from the exploratory calculations reported here and in
GNI07. For example, we would expect the gaseous disk to be puffed up
close to the star and to shield the outer part of the disk from
stellar radiation. On the other hand, the flaring of the outer disk
would also be increased, and thus it would be illuminated more
directly by the central star \citep{Dullemond2002,Gorti2004}. A
quantitative evaluation of such effects on the diagnostic line fluxes
must await the development of a more complete model that treats both
the gas and the dust self-consistently.

The D'Alessio model does not include X-rays. In our previous studies
based on this density distribution, we adopted a ``reference'' X-ray
luminosity of $L_X=2\times 10^{30}$~erg~s$^{-1}$ and a thermal
spectrum with $kT_X=1$~keV. It is well known that both the X-ray
luminosity and spectrum of young stellar objects can change with time,
and that they vary widely from source to source. The ionization rate
is not very sensitive to the temperature of the spectrum, as shown in
Fig. 5 of \citet{Igea1999}. In the {\it Chandra} Orion Ultradeep
Project (COUP) study of the Orion Nebula Cluster \citep{Getman2005},
the X-ray luminosity function for the entire cluster spans 5 decades
from $ \log L_X = 27-32$.  Two of these decades of variation may be
associated roughly with the dependence of $L_X$ on stellar mass and
one on a dependence on stellar age \citep{Feigelson2005}, leaving two
or more unexplained decades. Thus the X-ray luminosity of a
pre-main-sequence object with specific mass and age may vary over two
orders of magnitude. This is consistent with the findings of
\citet{Wolk2005} for a subset of 28 solar-mass stars in the Orion
Nebula Cluster with masses in the range $M_{*} = 0.9-1.2M_{\sun}$,
which have a median X-ray luminosity of $\log
L_X=30.25$~erg~s$^{-1}$. The spread of two orders of magnitudes can
only be partly explained by variability, including flares. Similar
results have been found in the XMM-{\it Newton} study of the
Taurus-Aurigae cluster \citep{Guedel2007, Telleschi2007}, where the
X-ray luminosity ranges over three decades ($\log L_X=28-31$).

Our previous choice $L_X=2\times 10^{30}$~erg~s$^{-1}$ was consistent
with the observations of young T Tauri stars with masses $M_{*} \sim 1
M_{\sun}$ and with the X-ray luminosity of well studied sources such
as TW Hya \citep{Kastner2002, Stelzer2004}. The median X-ray
luminosity measured by XEST for classical T Tauri stars (with disks)
in the Taurus-Aurigae cluster is $L_X=5\times 10^{29}$~erg~s$^{-1}$,
four times smaller than our reference value. In recognition of this
difference as well as in the observed 1-2 dex spread in measured
luminosity, we will consider a range of X-ray luminosities between
$L_X=2\times 10^{29}-2\times 10^{31}$~erg~s$^{-1}$. X-ray luminosities
in this range are likely to yield fluxes for the lines of interest
that can be detected with current facilities.

In order to improve the numerical accuracy of the flux calculations in
Section 4, the calculations reported here cover an extended range of
radii from 0.25-100\,AU, in contrast to GNI07 who considered the range
1-40\,AU. In making this extension, we do not consider any structural
and physical modifications that might occur at very small and very
large radii. For example, spectral energy distributions measured with
the {\it Spitzer} IRS indicate the occurrence of inner holes and rims
at distances of the order of several AU \citep{Dullemond2007}. Since
we use the D'Alessio model calculations, which varies continuously
from $0.028$ to $> 500$\,AU, the present calculations do not take into
account structures such as holes, gaps, and rims.

\begin{figure}
\centering
\includegraphics[angle=0,scale=.57]{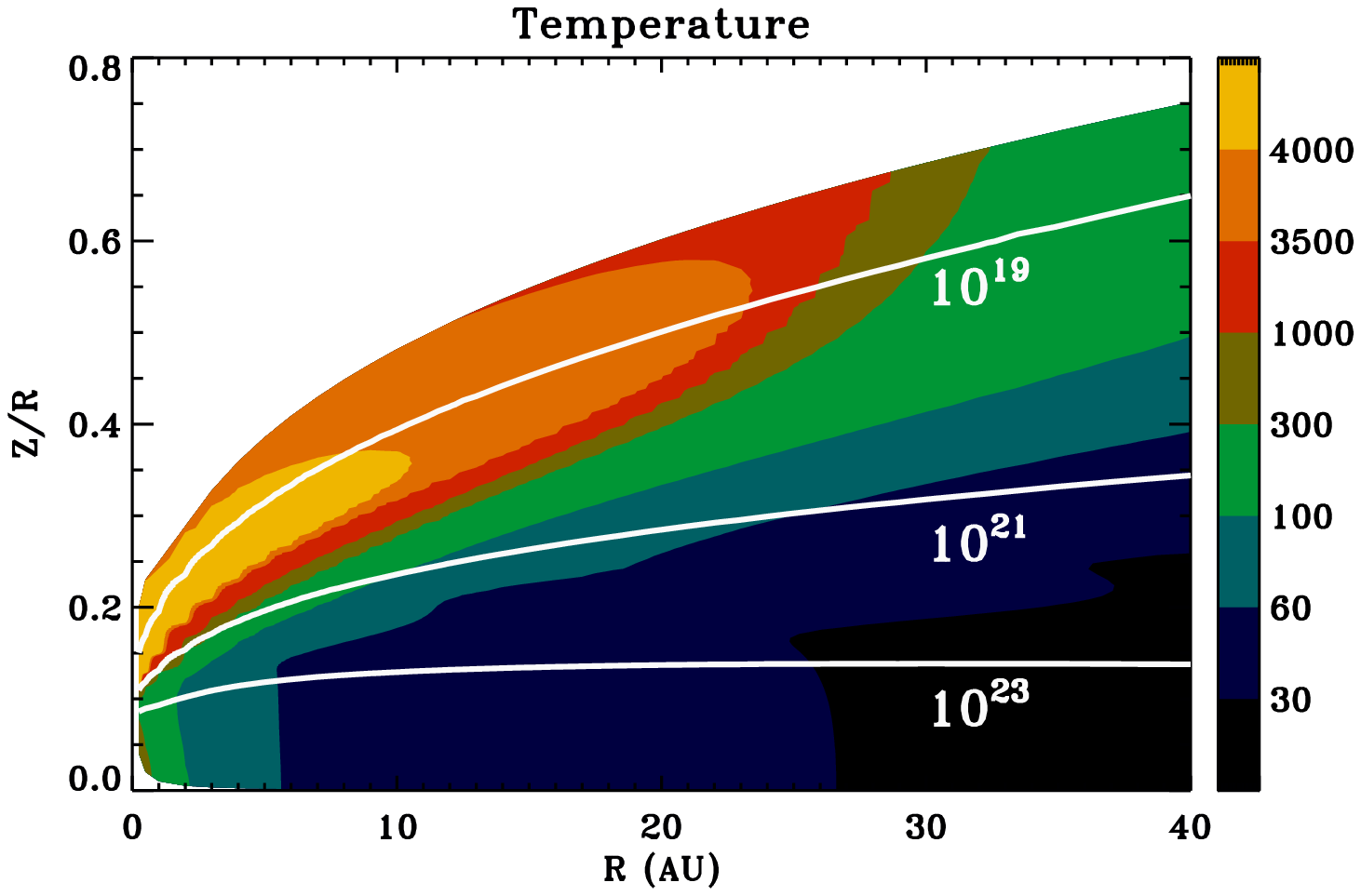}
\caption{Gas temperature structure of the reference model described in the
text with color code on the right. The density structure is based on
the \citet{dAlessio1999} model, defined by the parameters: $M_{*} =
0.5 M_{\sun}$, $R=2R_{\sun}$, $T_{*} = 4000$\,K, $\dot{M} =
10^{-8}\,M_{\sun}$\,yr$^{-1}$.  The reference X-ray luminosity is
$L_X=2\times 10^{30}$~erg~s$^{-1}$, and the temperature of the thermal
X-ray spectrum is $kT_X=1$\,keV.  The white curves are contours of
fixed vertical column density $N_{\rm H}$ in units of cm$^{-2}$.  The
curve for $N_{\rm H} = 10^{21}$\,cm$^{-2}$ corresponds roughly to two
scale heights as defined near the mid-plane. The integrated surface
density varies roughly as $1/R$ for $R > 1$\,AU and somewhat less
rapidly for $R < 1$\,AU.}
\label{temp}
\end{figure}

The elemental abundances used here and in the flux calculations to
follow are: $x_{\rm He}= 0.1$, $x_{\rm C} = 2.8 \times 10^{-4}$,
$x_{\rm O} = 6.0 \times 10^{-4}$, $x_{\rm S} = 7.0 \times 10^{-6} $,
$x_{\rm Ne}= 1.0 \times 10^{-4}$, and $x_{\rm Na} = 1.0 \times
10^{-6}$.  The abundance of sulfur that we have adopted is in the
range suggested by the recent modeling of sulfur molecules observed in
the Horsehead Nebula \citep{Goicoechea2006}. It is much closer to the
solar photospheric/meteoritic value \citep{Asplund2005} than the
heavily depleted values used in modeling dark clouds (e.g., Millar \&
Herbst~\citeyear{Millar1990}). UV absorption line studies of the
diffuse interstellar medium provide no evidence for depletion of
sulfur, as they do for more refractory elements
\citep{Savage1996}. Fig.~\ref{temp} shows the thermal structure of the
reference disk model.  The stellar and disk parameters that define the
D'Alessio 1999 model are given in the caption. For high altitudes
($N_{\rm H} < 10^{21}$~cm$^{-2}$) and moderate radii ($R<25$~AU),
temperatures as high as 4000-5000~K are reached.  Going deeper into
the atmosphere, $T$ undergoes a sharp drop toward mid-plane
temperatures in the range from 30 to 200\,K, depending on radius. Near
the mid-plane, where the density is high ($n=10^8-10^{10}$~cm$^{-3}$),
the gas and the dust are thermally coupled.  For these radii ($<
25$\,AU), the temperature starts high at the top of the atmosphere and
then increases further with increasing vertical column density or
decreasing altitude (see also Figure 2 of GNI04). This is a
consequence of the balance between X-ray heating and Ly-$\alpha$
cooling that controls the temperature at the very top of the
atmosphere.  The heating and the cooling are both proportional to the
first power of the volume density. The small increase in temperature
arises from {\bf its} logarithmic dependence on optical depth, which
of course increases with increasing vertical column density. Beyond
25\,AU, the O\,I fine-structure lines are the dominant coolants at the
disk surface.

\section{Ionization theory}

The GNI07 thermal-chemical program includes the elements H, He, C and
O, plus a generic heavy atom represented by Na. X-ray ionization of H,
He, H$_2$ and C are included explicitly, and that of O implicitly.
Generally speaking, X-ray ionization of a cosmic gas occurs by the
absorption by the K and L shell electrons of heavy atoms and ions. The
resultant photo and Auger electrons then generate many more secondary
electrons by the collisional ionization of H and He. The ionization
state of a heavy atom A or ion in a mainly atomic gas is determined by
several processes: direct X-ray ionization (rate $\zeta_{\rm dir}$);
secondary electron ionization (rate $\zeta_{\rm sec}$); electronic
recombination; and charge transfer to H and He. The latter process may
be fast (rate coeffcient $\sim 10^{-9}$\,cm$^{3}$s$^{-1}$) or slow. In
the case of neon (GNI07), charge transfer is relatively slow and
fosters high abundances of neon ions. The ionization rates for atom A,
$\zeta_{\rm dir}({\rm A})$ and $\zeta_{\rm sec}({\rm A})$, are always
per atom, whereas the total ionization rate $\zeta$ is per H nucleus.

On the basis of its elemental X-ray absorption cross section, we
estimate that the oxygen ionization rates are $\zeta_{\rm dir}({\rm
O}) = 17\zeta$ and $\zeta_{\rm sec}({\rm O}) = 2.4 \zeta$, where
$\zeta$ is the X-ray ionization rate of the cosmic mix of disk
gas. However, near-resonant charge exchange of H$^+$ and O is fast and
dominates the ionization of oxygen. Thus direct X-ray ionization of
oxygen can be ignored, as in GNI04. In the case of carbon, we
estimated $\zeta_{\rm dir}({\rm C}) = 6\zeta$ and $\zeta_{\rm
sec}({\rm C}) = 4\zeta$, but now charge exchange of H$^+$ and He$^+$
with C is very slow, and therefore X-ray ionization of carbon is
important and is included explicitly in the thermal-chemical program.
 
\begin{figure}
\centering
\includegraphics[angle=0,scale=.57]{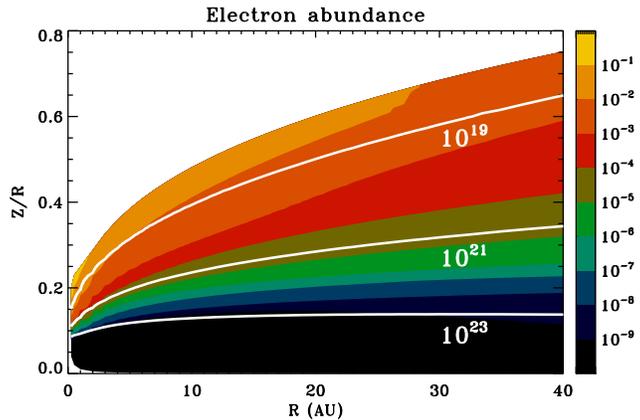}
\caption{Spatial distribution of the electron fraction plotted in the same 
way as Fig.~1, except for the color-coded units. \label{electron_abundances}}
\end{figure}

\begin{figure}
\centering
\includegraphics[angle=0,scale=.57]{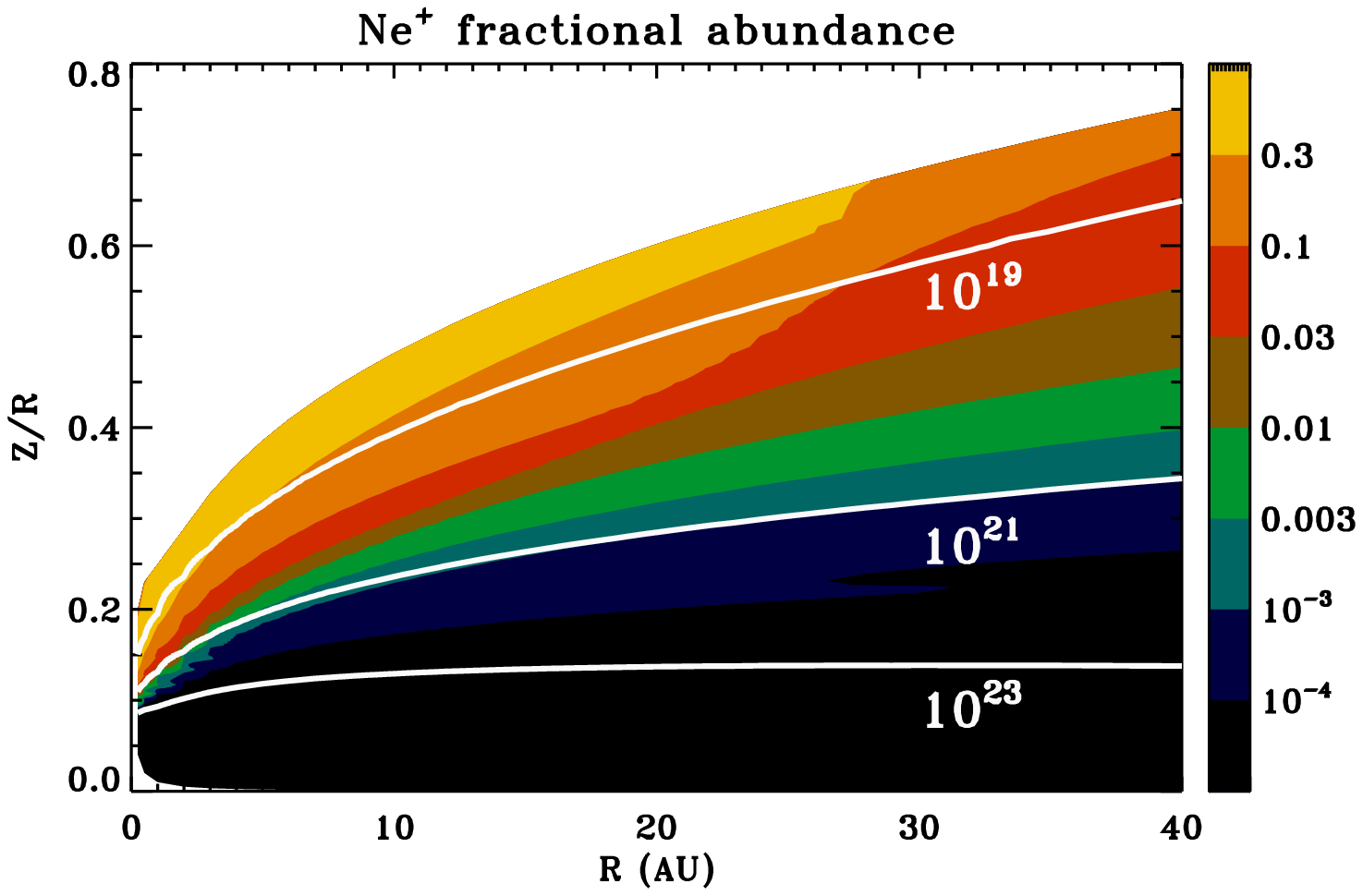}
\includegraphics[angle=0,scale=.57]{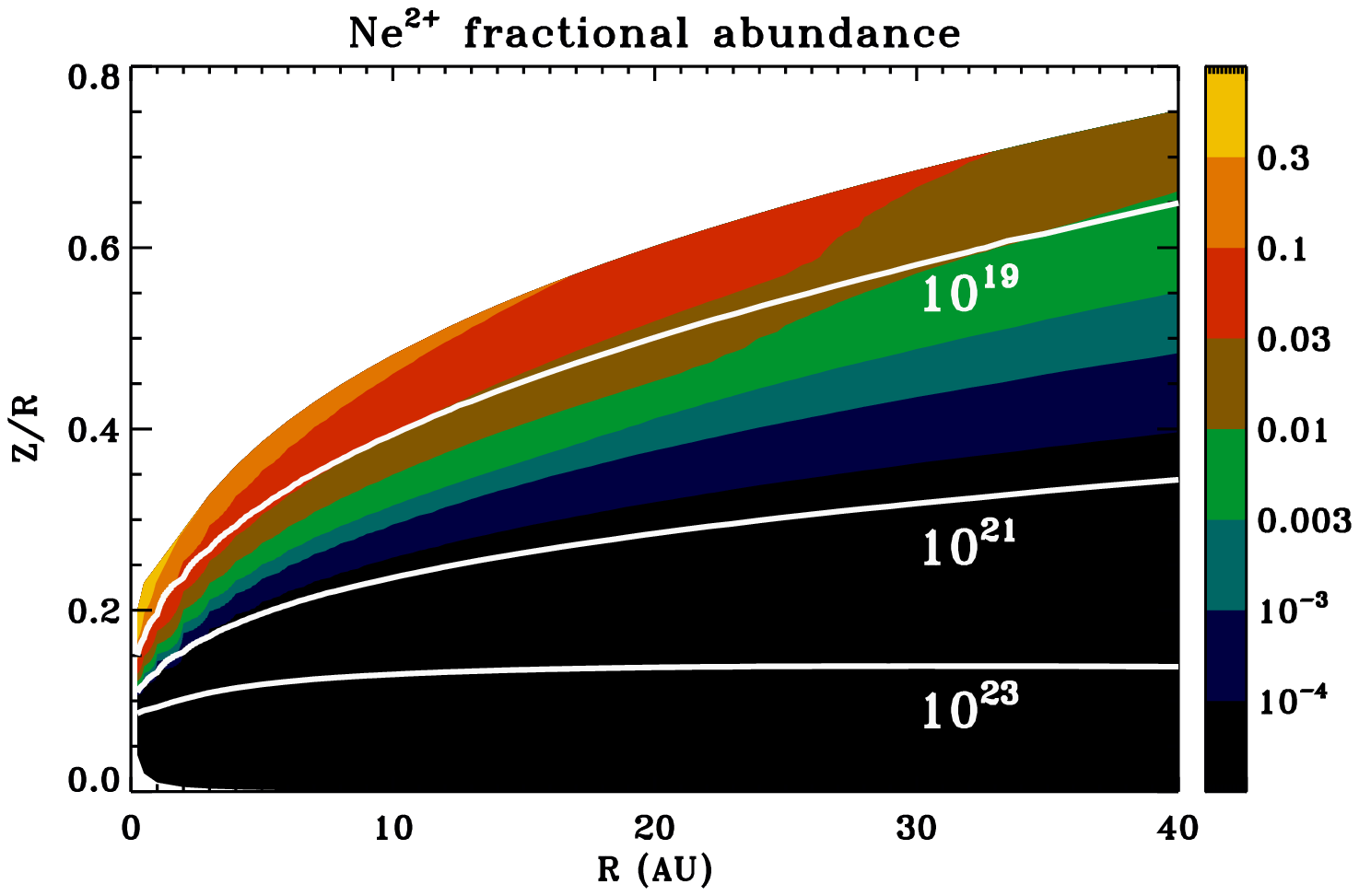}
\caption{Spatial distribution of the Ne$^+$ and Ne$^{2+}$ abundances 
relative to the total abundance of neon, plotted in the same way as Figure 1.}
\label{Neon_frac}
\end{figure}

Figure \ref{electron_abundances} shows the electron fraction
calculated for our model; it exceeds 0.01 at the top of the atmosphere
down to a depth of $N_{\rm H } \sim  10^{19}$cm$^{-2}$. The electron
fraction decreases more smoothly with vertical column than the
temperature. Near the mid-plane, its value is larger than that given
by GNI07 because ionization due to the radioactive decay of $^{26}$Al
have been included at a rate $\zeta_{26} = 4\times
10^{-19}$~s$^{-1}$ \citep{Stepinski1992, Glassgold1995,
Finocchi1997}. This assumes that the entire disk is optically thick
with respect to the 1.809\,MeV decay $\gamma$-rays, and that the
$^{26}$Al/$^{27}$Al ratio has the canonical value found in meteoritic
calcium-aluminum inclusions ($5\times 10^{-5}$). It also ignores the
effects of the 1\,Myr mean life of $^{26}$Al and the possibility that
the initial distribution of $^{26}$Al is spatially inhomogeneous. In
light of these simplifying assumptions, it is likely that the rate is
even smaller than the one we have adopted, but the exact value is not
crucial for the main subject of this work\footnote{Many modelers of
disk chemistry have used $^{26}$Al ionization rates an order of
magnitude larger than we have as the result of a typographical error
in the paper by \citet{Umebayashi1981}.}.  We use a separate
program to obtain the abundances of neon and sulfur species. The
ionization calculation of neon is essentially the same as
GNI07. Figure \ref{Neon_frac} shows the fractional abundances of
Ne$^+$ and Ne$^{2+}$ relative to the total abundance of neon.
Abundances of the order $\sim 0.1$ or more are found at high altitudes
($N_H < 10^{21}$~cm$^{-2}$). It is remarkable that X-ray ionization
produces significant abundances of Ne$^+$ and Ne$^{2+}$ out to radii
as large as 30\,AU.

\begin{figure}
\centering
\includegraphics[angle=0,scale=.44]{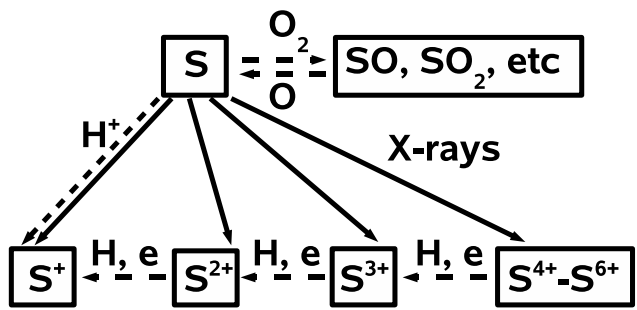}
\caption{Schematic diagram of sulfur chemistry. Charge exchange of 
sulfur ions with hydrogen atoms is critical for the ionization 
structure at high altitudes, whereas molecule formation controls 
the abundance of atomic sulfur near the mid-plane.} 
\label{sulfur_chemistry}
\end{figure}

The closed shell character of neon means that the complexities of
molecule formation and destruction do not enter, and only three
species, Ne, Ne$^+$ and Ne$^{2+}$ have to be considered in estimating
their line fluxes.  This is not the case for the calculation of the
S\,I and S\,II fine-structure and forbidden lines. In order to
calculate the abundances of S and S$^+$, we have to consider the
transition into sulfur molecules, such as SO, SO$_2$ and CS, which
occurs at large vertical column densities. Figure
\ref{sulfur_chemistry} gives a schematic overview of sulfur ionization
including a simplified warm chemistry, following \citet{Leen1988}.

Sulfur ions in the disk are mainly produced by X-ray ionization.
Sulfur has L and K edges near 0.2 and 2.5\,keV and, since we consider
X-ray spectra that are not particularly hard ($T_X \gtrsim 1$~keV),
photons with energies between the L and K edges are the ones most
strongly absorbed by sulfur. Because charge exchange with H is fast
for highly-ionized sulfur ions, X-ray ionization of sulfur accompanied
by the Auger process leads primarily to S$^{2+}$ and
S$^{3+}$. Calculations by \citet{Butler1980} and
\citet{Christensen1981} suggest that S$^{3+}$ charge exchange with
atomic hydrogen is also fast, while that of S$^{2+}$ is
slow. Therefore, we only include ions up to S$^{2+}$, as we did in the
case of neon (GNI07). However, in addition to electronic recombination
and charge exchange of sulfur ions with atomic hydrogen, we also
include the S + H$^+$ charge-exchange reaction, which can ionize S at
a significant rate below 10000~K according to the theoretical
calculations of \citet{Zhao2005}. Furthermore, the reaction S$^{2+}$ +
H$_2$, which has been measured to be fast \citep{Chen2003}, can
significantly reduce the abundance of S$^{2+}$.

As for carbon and oxygen, we can estimate the direct X-ray ionization
rate of sulfur from its X-ray absorption cross section. Its energy
dependence is very similar to that of the mean X-ray absorption cross
section averaged over interstellar or solar abundances. Thus the
direct ionization rate of sulfur, resulting from the (L to K)
0.2-2.4\,keV band is basically given by the ratio of the two
absorption cross sections at these energies.  For example, at 1\,keV,
the ratio is $\approx 500$. This then is the approximate ratio of the
primary or direct X-ray ionization of sulfur (per sulfur nucleus) to
that of the mean interstellar atom (per hydrogen nucleus, since
abundances are conventionally normed to hydrogen).  However, most of
the ionization of the gas as a whole comes from secondary-electron
ionization of hydrogen (atomic and molecular) and helium. Since there
are roughly 25 such electrons produced per primary ionization, the
ratio of the direct sulfur ionization rate per sulfur nucleus to the
total X-ray ionization per H nucleus is 500/25= 20:
\begin{equation}
\label{dirrate}
\zeta({\rm S})_{\rm dir} \simeq 20 \zeta.
\end{equation}

The secondary electrons generated by X-ray ionization of all of the
cosmic elements collisionally excite and ionize S at a rate that is
roughly given by the ratio of the electronic ionization cross section
of sulfur to that of hydrogen \citep{Maloney1996}. This leads to the
approximate value,
\begin{equation}
\label{secrate}
\zeta({\rm S})_{\rm sec} \simeq 5 \zeta.
\end{equation}
The direct ionization rate is mainly responsible for producing higher
S ions, which we lump together into ${\rm S}^{2+}$ because of fast
charge transfer. The secondary ionization rate primarily produces an
additional electron, e.g., it generates ${\rm S}^+$ from S.  Some of
the sulfur ion rate coefficients are listed in Table \ref{rates}.
Those for radiative recombination are well calculated. As in the case
of neon, however, the rate coefficients for electron transfer from atomic
hydrogen are not well established. They represent an important
uncertainty in the results of this paper.

To address the question of whether S remains the dominant species near
the mid-plane, we include the formation and destruction of the sulfur
molecules, SO, SO$_2$, CS and OCS, by using the warm neutral sulfur
chemistry of \citep{Leen1988}.  We also include X-ray ionization and
destruction of these molecules, assuming that X-ray cross sections for
molecules are additive over the constituent atoms, and using the
atomic rates discussed above.  We also make the simplifying
assumptions that a direct absorption of an X-ray by a sulfur molecule
always leads to dissociation, and that secondary electronic ionization
always produces a molecular ion. The sulfur molecular ions are rapidly
destroyed by dissociative recombination. The branching ratios are
summarized by \citet{Florescu2006}.

\begin{figure}
\centering
\includegraphics[angle=0,scale=.47]{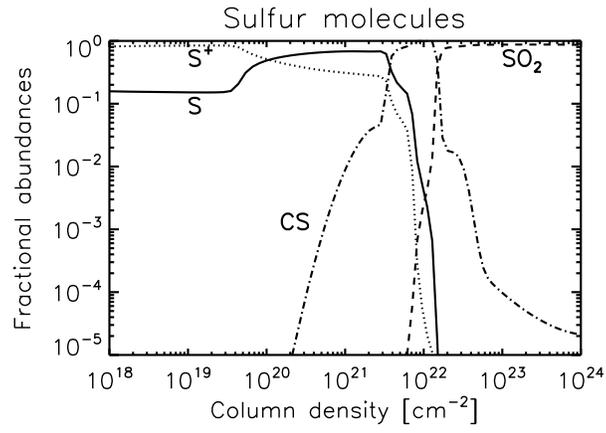}
\caption{Fractional sulfur abundances vs.~vertical column density 
measured from the top of disk at a radial distance of 20 AU from the
star. SO$_2$ becomes the dominant species at large columns.}
\label{sulfur_molecules}
\end{figure}

Figure \ref{sulfur_molecules} shows the fractional abundances of the
sulfur species S$^+$, S, SO$_2$ and CS as a function of vertical
column density above the disk mid-plane at a radial distance of 20\,AU
from the star. We find that sulfur is completely locked up in
molecules for column densities larger than
$0.5-1.0\times10^{22}$~cm$^{-2}$. In this region, neutral formation
reactions start to dominate over X-ray destruction as the X-rays get
shielded and the increased density enhances the reaction rates. The
main species at large column densities is SO$_2$, but there may also
be a layer of CS at intermediate column densities.

\begin{figure}
\centering
\includegraphics[angle=0,scale=.57]{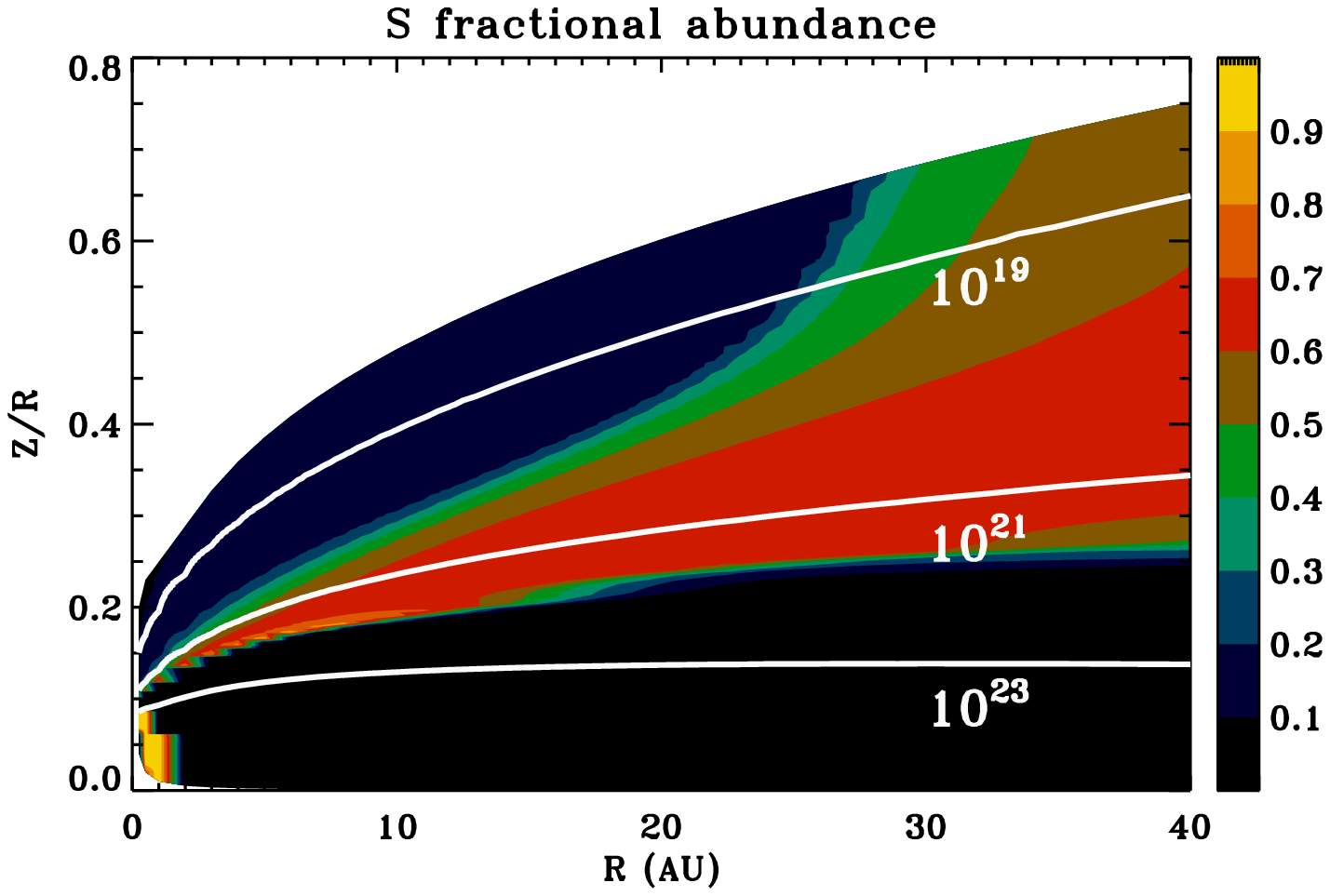}
\includegraphics[angle=0,scale=.57]{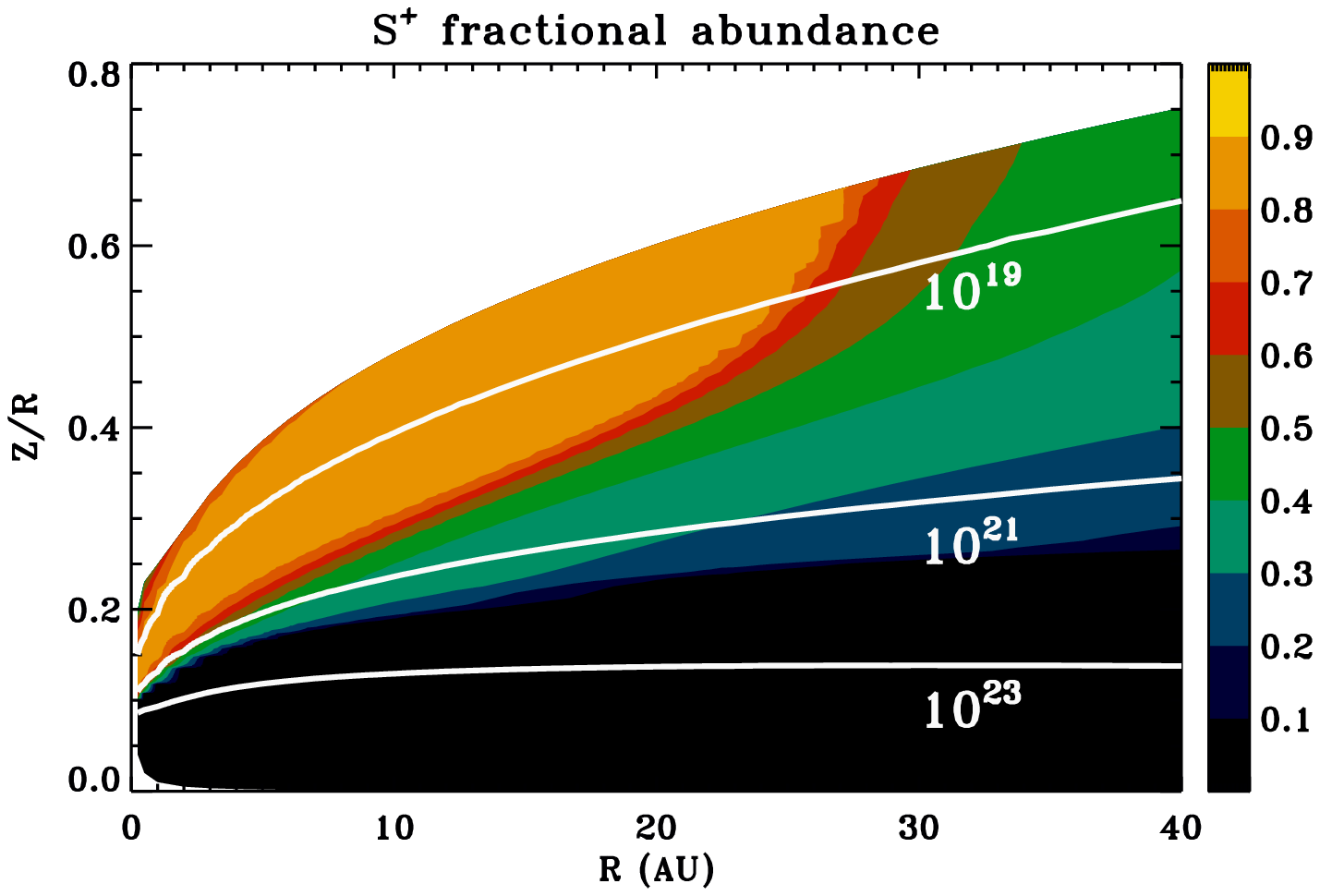}
\caption{Spatial distribution of normalized S and S$^{+}$ abundances,
plotted in the manner of Figure 1.} 
\label{S_fraction}
\end{figure}

Figure \ref{S_fraction} shows the S and S$^+$ fractions relative to
the total abundance of sulfur in the disk for radii between 0.25 and
40\,AU. S$^+$ is the dominant species at the highest altitudes for $R
< 25$\,AU , whereas S assumes that role at intermediate altitudes.
Beyond $R>25$~AU, X-ray ionization of sulfur is not strong enough to
maintain S$^+$ as the dominant species. At low altitudes, we find a
clear cut-off in the atomic S abundance, with almost all of the sulfur
in molecules.

\section{Excitation and Flux Calculations}

The flux calculations for the atomic ions considered in this paper are
relatively straightforward because one needs to deal only with a few
levels of relatively low excitation, even for the forbidden lines. For
neon, we continue to focus on the ground fine-structure transitions of
Ne$^+$ and Ne$^{2+}$. For O, S and C, we treat the 5 lowest levels in
the ground electronic configuration: a fine-structure triplet, that
produces mid and far IR lines, and two higher-energy singlet levels
that give rise to forbidden optical-IR transitions. For C$^+$, we
consider only the ground level fine-structure doublet that generates
the famous 158\,$\mu$m line. Simplifications often occur. For example,
the fine-structure transitions of C\,I and C\,II have low critical
densities and, because of the high disk densities, are
near-thermalized. Similarly, the forbidden lines are usually optically
thin and the effects of line trapping are small. In the following
sections, we discuss specific issues that arise in the calculation of
the emissivities of the lines of interest.  We use the line
frequencies and $A$-values in the NIST data base
(http://physics.nist.gov/PhysRefData/ASD/index.html). Electronic
excitation rates are now reasonably well established, but those for
excitation by H, H$_2$, He and H$^+$ are often unknown, except perhaps
for O\,I, C\,I and C\,II. References for our choice of rate
coefficients are given in Table \ref{excitation_refs}. In those cases
where trapping plays a role, e.g., for the O\,I and S\,I
fine-structure lines, our results apply to a face-on disk.

\subsection{Neon}

We calculate the optically-thin Ne\,II and Ne\,III lines fine-structure
lines at 12.81 and 15.55\,$\mu$m in the manner of GNI07 with the
following change. While retaining the two-level population formula for
the ground-level doublet of Ne\,II, we solve exactly for the populations
of the ground-level triplet of Ne\,III\footnote{GNI07 omitted the weak
quadrupolar transition that connects the top ($J=0$) and bottom
($J=2$) levels. Their approximate equation (3-4) also contained a
typo; the correct formula is,
\begin{equation}
P_u(2) = \frac{1.0}{1 + 5/3 C_{1-2} \exp{925.3/T} + 1/3 \exp(-399/T) / C_{0-1}}
\end{equation}}.
This more complete excitation calculation leads to an increase in the
population of the upper $J=0$ level and, to a lesser extent, the $J=1$
level, and increases the line intensities.

\begin{figure}
\centering
\includegraphics[angle=0,scale=.57]{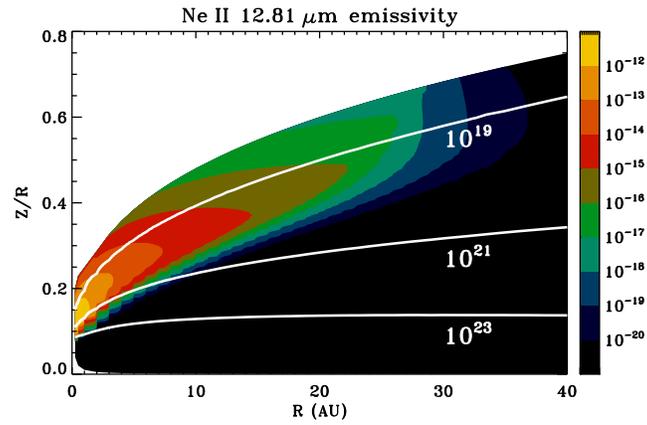}
\caption{The spatial variation of the Ne\,II 12.81\,$\mu$m fine
structure emissivity, defined by equation (1), in units of erg
cm$^{-3}$s$^{-1}$.}
\label{Neon_emission}
\end{figure}

Figure \ref{Neon_emission} shows the distribution of the Ne\,II
12.81\,$\mu$m emissivity throughout the disk for the reference X-ray
luminosity. Most of the emission is produced at high temperature
($\sim 4000$~K) and high electron abundance, corresponding to radial
distances $R < 25$~AU and vertical column densities
$N_H<10^{21}$~cm$^{-2}$. The largest emissivity does not always occur
at the highest altitude above the mid-plane. This is due to an
increase of the Ne$^{+}$ density and its specific emissivity with
increasing depth. The upper $J=1/2$ level is sub-thermally populated
and its population is proportional to the electron density, which also
increases going down into the atmosphere.

The above calculations assume that the excitation is solely due to
electronic collisions. However, we have been reminded that, as the
electron fraction decreases with height, collisions with atomic
hydrogen begin to play a role (D. J. Hollenbach 2006, private
communication). Using the results of \citet{Bahcall1968},
\citet{Hollenbach1989} estimated the rate coefficient for the
de-excitation of the 12.81\,$\mu$m transition in collisions with
atomic hydrogen to be $1.3\times10^{-9}$~cm$^{3}$~s$^{-1}$. Although
the theory of Bahcall and Wolf may give the right order of magnitude
(as it does for the excitation of the O\,I fine-structure
transitions), it is incorrect in principle since it implies that the
excitation cross section is proportional to the scattering cross
section. According to Bahcall and Wolf, the rate coefficient in a
collision with a neutral atom varies with temperature as $T^{1/6}$
(c.f.~the van der Waals potential) and as a constant in a collision
with an ion (c.f.~the Langevin potential). However, collisional
excitation of a fine-structure transition requires an exchange of a
spin between the incident H atom and the target electrons and thus
depends on differences between potentials, differences which decrease
more rapidly than $1/r^6$ and $1/r^4$ for neutral and ionic
collisions, respectively.

The full quantum-mechanical calculations for O\,I and C\,I
\citep{Launay1977a, Abrahamsson2007} and for C\,II \citep{Launay1977b,
Barinovs2005} bear out the fact that the Bahcall and Wolf theory
predicts the wrong temperature dependence for the collisional
excitation rate coefficients by atomic hydrogen.  We have compared the
recent calculations of \citet{Barinovs2005} for H + C\,II with the
Bahcall and Wolf theory, and find that the latter over-estimates the
rate coefficient by factors of 7.5 to 5 in the temperature range from
500-4000\,K. If something like this reduction factor also applies to H
+ Ne\,II, then the order of magnitude of its rate coefficient for
collisional de-excitation of the Ne\,II fine-structure doublet is
$\sim 2\times 10^{-10}$cm$^{3}$s$^{-1}$.  We have calculated the
12.81\,$\mu$m emissivity (including line trapping) with this value for
the H de-excitation rate, and find that it increases the flux by about
a factor of two. This increase arises mainly at large radial
distances, where the deviations from a thermal population are the
greatest. We conclude that our calculations based on electronic
excitation alone underestimate the Ne\,II 12.81 $\mu$m flux, but
possibly not by a large factor. Actual calculations of the atomic
hydrogen excitation rate would be most welcome.

\subsection{Oxygen}

\begin{figure}
\centering
\includegraphics[angle=0,scale=.44]{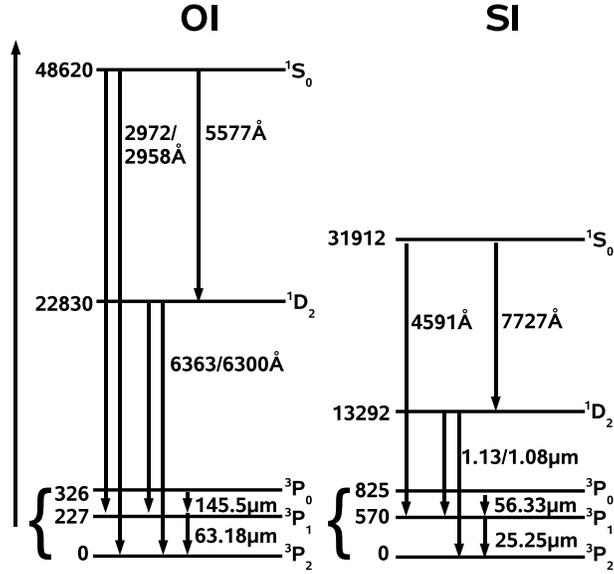}
\caption{Energy level diagrams for O and S showing the low-lying levels that
generate the fine-structure and forbidden lines. Note that the energy scale 
is in Kelvin.}
\label{excitation_diagram}
\end{figure}

Figure \ref{excitation_diagram} gives the 5 levels that we consider in
calculating fine-structure and forbidden line emission from atomic
oxygen and sulfur.  According to Table \ref{excitation_refs},
collisional excitation rates for the fine-structure levels of O\,I are
available for collision partners e, p, H, H$_2$ and He. Atomic
hydrogen tends to dominate and, even without line trapping, the
critical densities are modest, typically $n_{\rm cr}({\rm H}) <
10^6$cm$^{-3}$. At radii ($>25$\,AU), densities of this order are not
reached until vertical column densities $N_{\rm H} \approx
10^{20}-10^{21}$~cm$^{-2}$. The full emissivity calculation is
particularly relevant at large radii.

\begin{figure}
\centering
\includegraphics[angle=0,scale=.57]{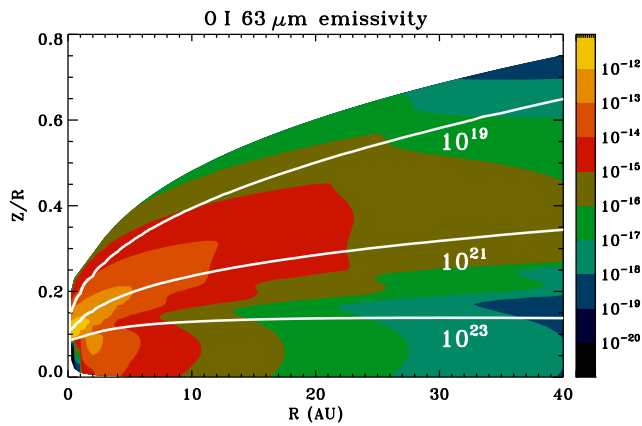}
\caption{Spatial distribution of the O\,I 63\,$\mu$m fine-structure
line emissivity, defined by equation (1), in units of erg
cm$^{-3}$~s$^{-1}$.}
\label{OI_finestruct}
\end{figure}

\begin{figure}
\centering
\includegraphics[angle=0,scale=.57]{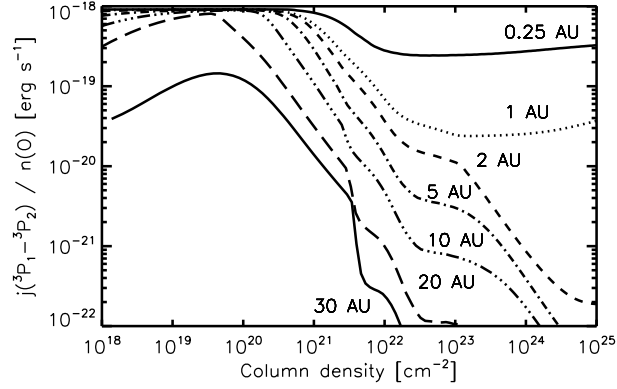}
\caption{The specific O\,I 63\,$\mu$m emissivity in units of 
erg s$^{-1}$ atom$^{-1}$, vs. perpendicular column density at
various radii.}
\label{oxygen_yield}
\end{figure}

Figure \ref{OI_finestruct} shows the spatial distribution of the
emissivity of the O\,I 63\,$\mu$m fine-structure line. The results for
the O\,I 146\,$\mu$m line (not shown) are similar. The spatial profile
is particularly complicated at small radii ($<10$\,AU).  The
emissivity has two peaks going down into the disk (increasing vertical
column at fixed radius) that result from a combination of effects:
density and temperature variations, abundance changes and
line-trapping. The declining temperature, the conversion of atomic
oxygen into molecules (CO, H$_2$O and O$_2$), and trapping at large
columns all tend to decrease the emissivity. These effects are
illustrated by Figure \ref{oxygen_yield}, which shows the {\it
specific emissivity} (per O atom). For small column densities, it is
constant when the population is thermalized (at radii $R < 10$~AU,
where the critical densities are reached), or it increases toward a
maximum when the critical densities are only attained at larger
vertical column densities. Then it decreases again with vertical
column density, except at the smallest radii ($<1$\,AU), where the
rise in temperature due to viscous heating close to the mid-plane has
the opposite effect. The occurrence of the two peaks in emissivity at
intermediate columns is the result of the rapid increase of volumetric
density with increasing depth. Significant emission occurs beyond
columns of $N_{\rm H} = 10^{22}$\,cm$^{-2}$ as a consequence of the
fact that the excitation energy is relatively low ($\approx 230$\,K)
and contributes $\sim10$\% of the total line flux.

The $^1D_2-^3P_J$ O\,I 6300/6363\,\AA\ and the $^1S_0-^1D_2$
5577\,\AA\ forbidden transitions are optically thin and sub-thermally
excited.  In addition to the electronic rate coefficients
\citep{Zatsarinny2003}, the collisional de-excitation rate of the
$^1D_2$ level by atomic hydrogen has been calculated by
\citet{Krems2006}. It varies little between 500-6000\,K, where it has
a mean value of $8\times 10^{-13}$~cm$^{3}$~s$^{-1}$ to within 25\%.
At 4000\,K, the collisional electron de-excitation rate is $3\times
10^{-9}$~cm$^{3}$~s$^{-1}$ for the $^1D_2$ level, so electronic
collisions dominate over those with atomic hydrogen for $x_e>2.5\times
10^{-4}$. If we consider only electronic collisions, then an upper
limit to the critical density for the 6300/6363\,\AA\ transitions is
$n_e \approx10^6$~cm$^{-3}$, while that for the 5577\,\AA\ transition
is $n_e \approx 10^8$~cm$^{-3}$.  In the absence of calculations for
the $^1S_0$ level, we have adopted the value,
$10^{-12}$~cm$^{3}$~s$^{-1}$, for the H rate coefficient. Thus the
6300 and 6363\,\AA\ lines are populated near-thermally, and we can use
the two-level formulae given in equations (3-2) and (3-3) of GNI07
that allow for the effects of finite values of the critical
density. The emissivity for both lines is then given by
equation~\ref{thermalemission}. The $^1S_0$ uppermost level is so far
from being thermalized that we can calculate the flux of the
5577\,\AA\ line by using the extreme low-density
collisional-excitation limit, in which every excitation leads to the
production of a photon.

\begin{figure}
\centering
\includegraphics[angle=0,scale=.57]{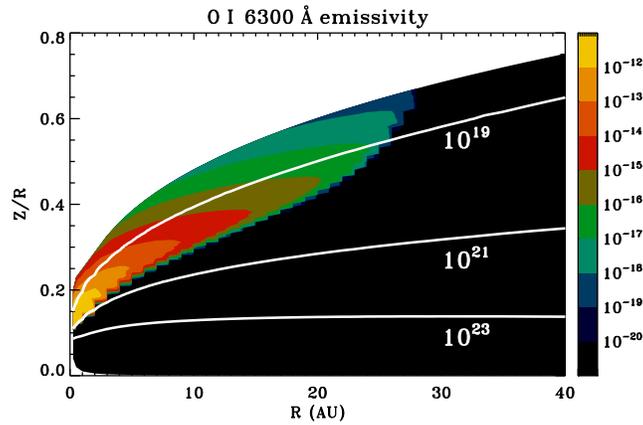}
\caption{Spatial distribution of the O\,I 6300\,\AA\ forbidden line
emissivity, in units of erg cm$^{-3}$s$^{-1}$.}
\label{OI_forbidden}
\end{figure}

Figure \ref{OI_forbidden} shows that the O\,I forbidden lines trace
the upper atmosphere of the disk, where the atomic oxygen abundance is
constant. The O\,I 6300/6363\,\AA\ lines are generated within 25\,AU,
a somewhat smaller range than the neon fine-structure lines, while the
5577\,\AA\ line (not shown) originates from radii inside 15\,AU.
These calculations of forbidden line fluxes necessarily ignore an
effect of the UV photodissocation of OH, which leads to significant
branching to the exited $^1D_2$ level of O, producing fluorescent
emission of the O\,I 6300/6363\,\AA\ lines
\citep{Stoerzer2000}. Because significant OH abundances only occur at
large depths in our model, this effect might become important at large
radii due to OH photo-dissociation by the stellar or
interstellar/intracluster FUV radiation field.

\subsection{Sulfur}

We focus on the mid-IR fine-structure lines of S\,I and the
6718/6732\,\AA\ forbidden transitions of S\,II; the fine-structure
lines of S\,III are likely to be too weak to be observable at the
present time. In accord with Figure~\ref{excitation_diagram}, the
calculation of the S\,I fine-structure emission is similar to that
just described for O\,I.  The big difference is that little if
anything is known about the excitation of these lines by collision
partners other than electrons. Thus the present flux estimates are
lower limits.  The calculations take line trapping into account and
solve the three-level fine-structure population problem exactly,
including the the weak $J=0-2$ quadrupolar transition.

\begin{figure}
\centering
\includegraphics[angle=0,scale=.57]{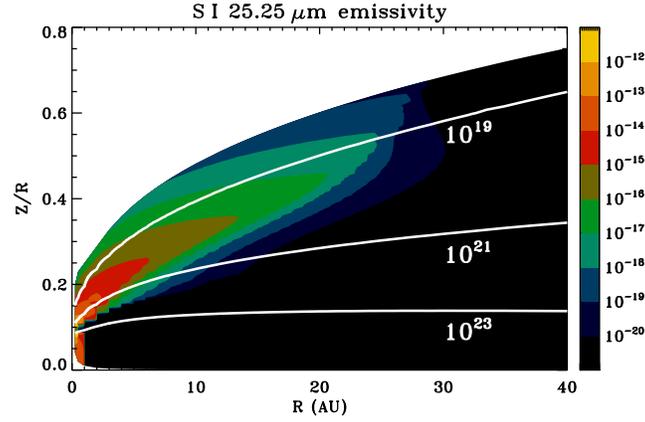}
\caption{Spatial distribution of the S\,I 25.55\,$\mu$m fine-structure
line emission in units of erg cm$^{-3}$ s$^{-1}$. The emission peaks
at intermediate depths, as discussed in the text.}
\label{SI_finestruct}
\end{figure}

\begin{figure}
\centering
\includegraphics[angle=0,scale=.57]{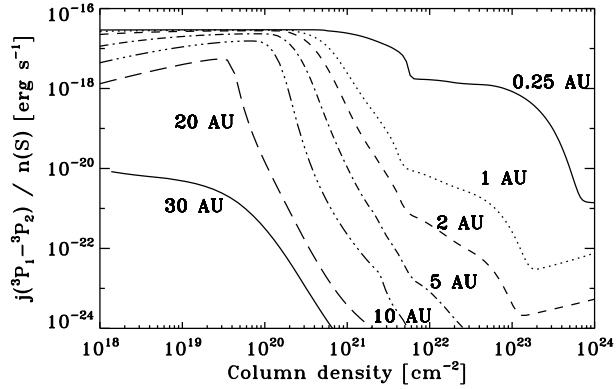}
\caption{The spatial distribution of the specific emissivity in units of
erg cm$^{-3}$ s$^{-1}$ atom$^{-1}$ of the S\,I 25.55\,$\mu$m fine-structure 
line vs. perpendicular column density at several radii.}
\label{sulfur_yield}
\end{figure}

Figure \ref{SI_finestruct} shows the spatial distribution of the S\,I
25.25\,$\mu$m fine-structure line emissivity for the reference model.
The appearance of the 56.33\,$\mu$m emissivity (not shown) is similar,
but it is concentrated toward smaller radii and higher altitudes.  As
in the case of O\,I, the variation of the fine-structure emission with
depth at fixed radius is non-monotonic, and again there are several
factors that contribute to this behavior. First, there are the
chemical changes already mentioned, where sulfur changes from S$^+$ to
S at high altitudes and from S to sulfur molecules at low
altitudes. Thus the emission peaks at intermediate altitudes, at a
vertical column of $N_{\rm H}\sim 10^{21}$~cm$^{-2}$ at $R=0.25$~AU
and near $N_{\rm H}\sim 10^{19}$~cm$^{-2}$ at $R=25$~AU.  Second, the
increase of volumetric density with depth also promotes fine-structure
emission, whereas the overall temperature decrease has the opposite
effect.  Figure~\ref{sulfur_yield} shows the variation of the specific
emissivity with vertical column at fixed radius. For radii in the
range 1-25\,AU, the emissivity per S atom of the $J=1-0$ level changes
from sub-thermal to thermal as the density increases with increasing
depth. At somewhat larger column densities, the specific emissivity
decreases due to decreases in both temperature and electron density.

The transition of atomic sulfur into molecules such as SO and SO$_2$
has little effect on the absolute level of the S\,I fine-structure
emission. This is due to the fact that electronic excitation has
already become ineffective for $N_{\rm H} < 10^{22}$\,cm$^{-2}$ before
molecule formation occurs. In order to gauge the importance of H atom
collisions, we carried out a calculation where we assumed that the
S\,I fine-structure de-excitation rates are about the same as for
O\,I, specifically $k({\rm H}) = 5.0\times 10^{-11}\ T^{0.4}$ for all
transitions.  The S\,I 25.25 and 56.33\,$\mu$m emission is then
increased by factors of 10 or more over that for electron collisions
only.  This serves to indicate the importance of extending the
calculation of H excitation rates to heavy atoms and ions such as
Ne\,II and S\,I.

\begin{figure}
\centering
\includegraphics[angle=0,scale=.57]{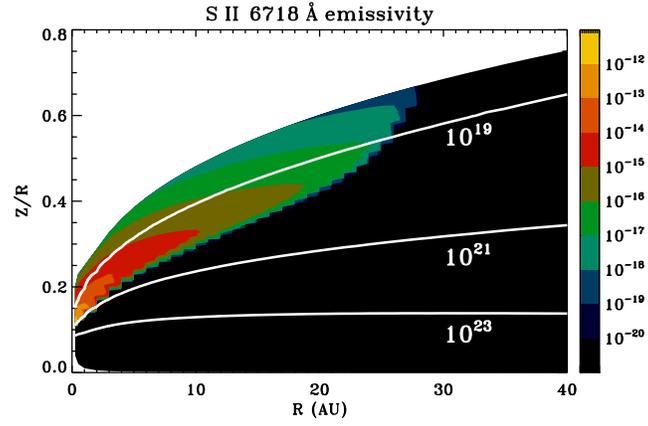}
\caption{Spatial distribution of the S\,II 6718\,\AA\ forbidden line emission
in units of erg cm$^{-3}$ s$^{-1}$.
Significant emission only occurs in the warmest part of the disk atmosphere.}
\label{SII_forbidden}
\end{figure}

We estimate the emission of the S\,II forbidden lines in the optically
thin approximation. We concentrate on the $^2D_{5/2,3/2}$ -
$^4S_{3/2}$ transitions with wavelengths near 6700\,\AA.  Using the
electron collisions strengths from \citet{Tayal1997}, we find electron
critical densities, $n_{\rm cr} \sim 10^4 -
10^5$\,cm$^{-3}$\,s$^{-1}$, and conclude that the lines are almost
thermalized.  We also considered the possible role of the H$^+$ + S
$\rightarrow$ S$^+$ + H charge exchange, which mainly branches to the
upper $^2P_J$ level of S$^+$, and then decays with the emission of
photons with wavelengths near 4070\,\AA, 6700\,\AA, and 1\,$\mu$m. We
find that this process does not increase the 6718 and 6732\,\AA\ line
emissivities, probably because the rate coefficient for charge
exchange is small ($\sim 5 \times 10^{-12}$~cm$^{3}$~s$^{-1}$) and
because the high density tends to a thermalize the level
population. In Figure \ref{SII_forbidden}, we show the emissivity of
S\,II 6718\,\AA\ line (the 6733\,\AA\ line is very similar).  These
lines trace the warm upper layers of the disk. Significant emission
arises only from regions with temperatures in the range 2000-5000\,K.

\subsection{Carbon}

The energy levels of the ground configuration of C\,I are similar to
those of O\,I and S\,I shown in Figure ~\ref{excitation_diagram},
except that the scale of the energy level separations is reduced and
the ordering of the total angular momentum quantum number $J$ in the
ground state triplet is inverted.  The upper fine-structure levels
have energies $E_2/k = 62.5$\,K and $E_1/k = 23.62$\,K and give rise
to far-infrared lines with wavelengths $\lambda(1-0) = 609.135\,\mu$m
and $\lambda(2-1) = 370.144\,\mu$m. In addition to these lines, we
calculate the emissivity of the forbidden lines at 9827 and 9853\,\AA\
that emanate from the first $^1D_2$ level of C\,I above ground. For
C\,II, we only consider the fine-structure doublet, with excitation
energy and wavelength, $\Delta E(3/2-1/2)/k = 91.2$\,K and
$\lambda(3/2-1/2) = 157.7$~$\mu$m, and ignore the 2300\,\AA\ lines
that arise from the next $^4P$ level.

The $A$ values of the C\,I and C\,II fine-structure transitions are
very small ($\sim 10^{-7}$\,s$^{-1}$ for C\,I and $2.3 \times
10^{-6}$\,s$^{-1}$ for C\,II), which implies that the critical
densities for all of these transitions are low ($<10^4$~cm$^{-3}$,
based on the collisional rate coefficients referenced in Table 2).
Thus we can estimate the emissivities assuming that the fine-structure
levels are in thermal equilibrium (but we do include
line-trapping). The integrated fluxes for a nominal distance of
140\,pc are given in Table 3. The flux of the C\,II 158\,$\mu$m line
is very small, but the 369\,$\mu$m line of C\,I appears to be in the
observable range, as may even be the case for the 609\,$\mu$m line.

\begin{figure}
\centering
\includegraphics[angle=0,scale=.57]{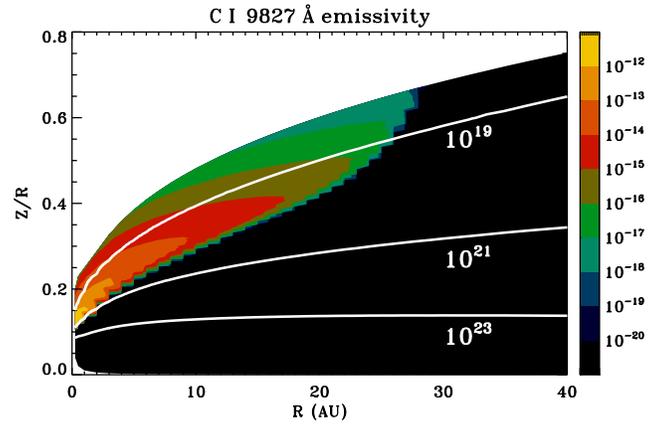}
\caption{Spatial distribution of the C\,I forbidden line emission in 
units of erg cm$^{-3}$ s$^{-1}$.}
\label{CI_forbidden}
\end{figure}

We estimate the the C\,I 9827 and 9853\,\AA\ forbidden line emission
in the optically thin approximation. Using Table 2, the critical
density of the $^1D_2$ level at 4000 K is
$1.96\times10^4$~cm$^{-3}$~s$^{-1}$, and the level is close to being
thermalized. The density-dependent correction for sub-thermal
excitation is made in the same way as for O\,I, as described in
Section 4.2. The spatial distribution of the C\,I 9827\,\AA\ emission
is shown in Fig. \ref{CI_forbidden}. The emission is peaked at small
vertical column densities, much like the O\,I 6300\,\AA\ emission in
Fig.~\ref{OI_forbidden}. The integrated line flux given in Table 3 is
somewhat smaller than the flux of the O\,I 6300\,\AA\ line, but still
in the potentially observable range. Of course all of these statements
about the detectability of neutral carbon lines stand to be corrected
(and reduced) when interstellar/intracluster UV radiation is included
in the ionization theory.

\section{Discussion}

In this section we discuss the most important results of this paper in
the context of both existing and prospective observations.  We treat
high and low excitation lines separately, since they trace different
parts of the disk. We first summarize in Table 3 the line fluxes for
five values of the stellar X-ray luminosity that range from
$L_X=2\times 10^{29} - 2\times 10^{31}$~erg~s$^{-1}$.  To ensure
accurate results for values of $L_X$ larger than standard, we have
calculated the disk properties and emissivities out to 100\,AU. The
calculations pertain to the case where X-ray heating of the disk
atmosphere dominates mechanical heating ($\alpha_h = 0.01$ in the
notation of GNI04). As discussed in Section 2, the underlying density
model is the generic T Tauri disk model of \citet{dAlessio1999}; the
parameters for the model are given in the caption of Fig.~1. The
``reference'' model that we have used throughout has an X-ray
luminosity of $L_X=2\times 10^{30}$~erg~s$^{-1}$; the fluxes for this
case are given in the middle column of the Table.

The values shown in the other columns illustrate how sensitive the
modeling results are to the choice of X-ray luminosity. Table 3
encompasses a range of 100 in X-ray luminosity, which is roughly the
spread seen by the COUP and XEST projects for T Tauri stars in the
Orion Nebula and Taurus-Aurigae clusters of the same mass and age.
Were we to choose the median $L_X$ for Taurus-Aurigae
\citep{Telleschi2007} for the reference model, the second column of
fluxes would be appropriate.  According to the discussion in the
previous sections, all of the flux estimates are themselves uncertain
to varying degrees due to uncertainties in the rate coefficients and
other limitations of the model.  In particular, the fluxes for the
neon and sulfur fine-structure lines are lower limits because of the
omission of collisional excitation by H atoms.

\begin{figure}
\centering
\includegraphics[angle=0,scale=.57]{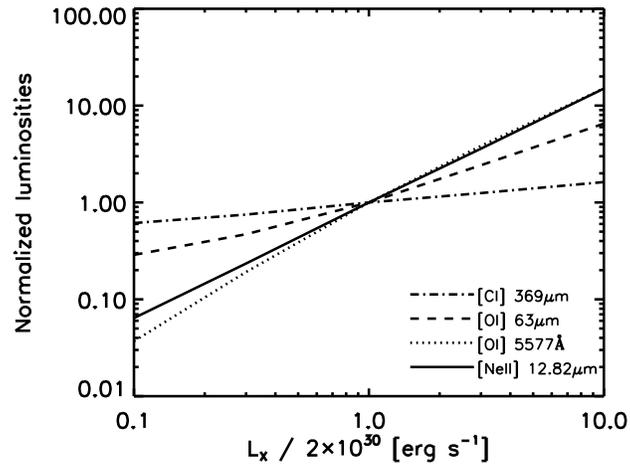}
\caption{The variation of selected line fluxes with X-ray luminosity, 
all normalized to the values for the reference model.}
\label{xray_trends}
\end{figure}

The sensitivity of the integrated line fluxes to the X-ray luminosity
can vary from one line to another, since they may originate in
different parts of the disk with different physical properties. We
illustrate this in Fig.~\ref{xray_trends}, which shows how some of the
line fluxes vary with $L_X$, normalized to the standard model. The
C\,I 369\,$\mu$m line luminosity depends very weakly on $L_X$, since
most of the emission is produced at intermediate column densities
($N_{\rm H}\sim 10^{22}$~cm$^{-2}$) where the X-ray flux is strongly
attenuated. The O\,I 63\,$\mu$m luminosity shows a somewhat steeper
dependence, because it is produced both at high altitudes, where the
X-ray flux is large, and close to the midplane, where the X-ray flux
is small (and the gas and dust temperatures are almost the same). The
optically thin O\,I 5577\,\AA\ emission displays the strongest
dependence on $L_X$ because, with its relatively high excitation
temperature, it is only produced in the warmest part of the disk fully
exposed to stellar X-rays. The other forbidden lines and the S\,I
fine-structure lines (not shown), behave very much like the Ne\,II
fine-structure line. The almost linear relation between Ne\,II line
flux and $L_X$ arises from the dependence of the emissivity on $n_e^2$
(c.f. Eq.~4-1 of GNI07). In its present form, Fig,~\ref{xray_trends}
is {\it not} meant to suggest a general empirical correlation of line
fluxes with X-ray luminosity, simply because the stellar and disk
properties of the reference model have been held fixed in the
calculations for this figure.

\subsection{High excitation lines}

\subsubsection{Neon Fine-Structure Lines}

\begin{figure*}
\centering
\includegraphics[angle=0,scale=.47]{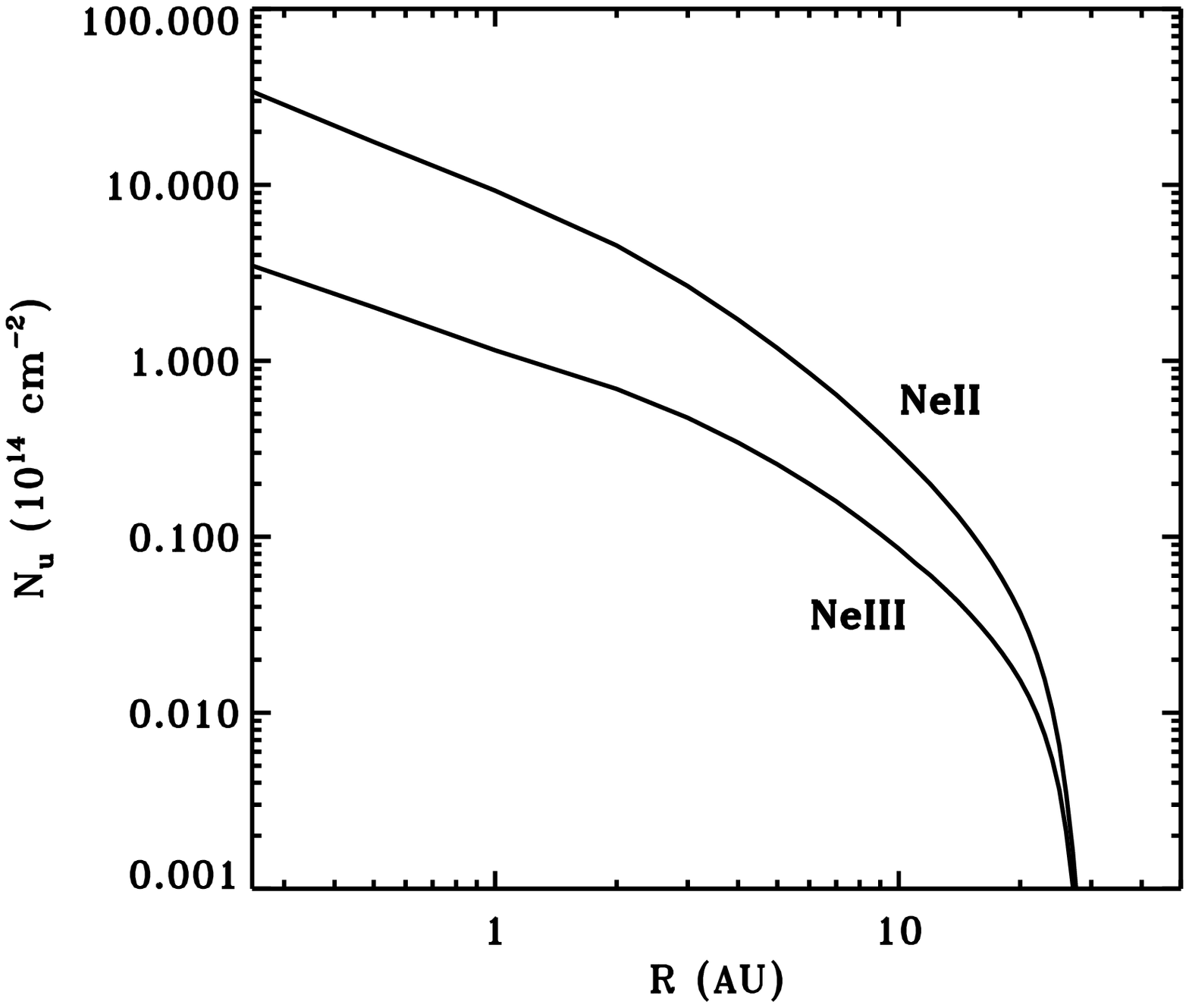}
\includegraphics[angle=0,scale=.47]{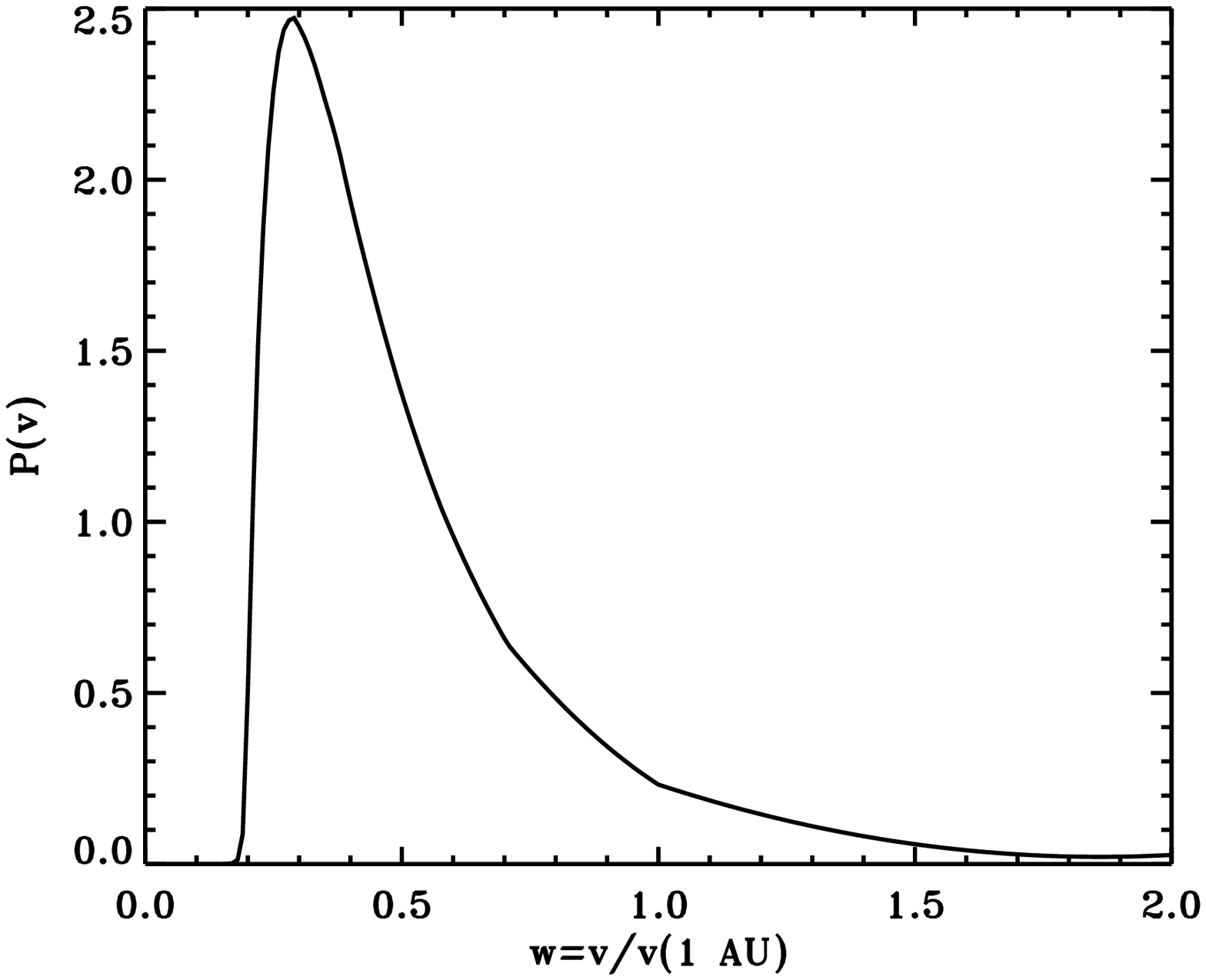}
\caption{Column densities of excited Ne ions in units of
$10^{14}$\,cm$^{-2}$ plotted vs.~radial distance in AU (left), and the
rotational velocity distribution function $P(v)$ plotted vs.~velocity
normalized to its value at 1\,AU (right).}
\label{neon_distribution}
\end{figure*}

The neon fine-structure line fluxes (Table
\ref{integrated_luminosities}) are close to but more accurate than
GNI07. The Ne\,III flux is 40\% larger since the calculations extend
to smaller radii ($R=0.25$~AU).  The Ne\,II 12.81\,$\mu$m line
emission arises from the top of the atmosphere, where the temperature
and electron density are high. We find significant contributions to
the integrated emission out to $R=25$~AU.  The emissivity per unit
radius is determined by the column density in the upper level
(c.f.~Eq.~3-7 of GNI07). This quantity is plotted in the upper panel
of Fig.~\ref{neon_distribution} against the radius $R$. In contrast to
Fig.~4 of GNI07, there is no dip near 5\,AU, thanks to more accurate
and closer spaced calculations. When we assume Keplerian rotation to
convert from column density to a velocity distribution function $P(v)$
(c.f.~Eq.~3-8 of GNI07), the peak in $P(v)$ occurs at $v=0.25v({\rm
1~AU})$, which corresponds to a peak in the emissivity per unit area
at 16\,AU. The bottom panel of Fig.~\ref{neon_distribution} shows that
$P(v)$ has a long tail extending to high velocity. The ratio of the
Ne\,III 15.55\,$\mu$m to the Ne\,II 12.81\,$\mu$m flux is of order
0.1. GNI07 considered the detection of the Ne\,III line to be a near
definitive diagnostic of X-ray ionization of neon. However, using a
disk model dominated by stellar EUV radiation, Hollenbach \& Gorti
(2007, private communication), find a range of possible NeIII/NeII
ratios (0.1-6) depending on the the nature of the EUV spectrum. Thus
it is not immediately clear whether the Ne\,III/Ne\,II fine-structure
line ratio can serve as a discriminant between EUV and X-ray
irradiation.

The Ne\,II 12.81\,$\mu$m line has now been detected in T Tauri disks
in about 20 cases from the ground and from space (by the IRS on {\it
Spitzer}: Pascucci et al.~\citeyear{Pascucci2007}; Lahuis et
al.~\citeyear{Lahuis2007}; Espaillat et al.~\citeyear{Espaillat2007};
and MICHELLE on Gemini North: Herczeg et al.~\citeyear{Herczeg2007}).
The detected flux levels are of order
$10^{-14}$~erg~cm$^{-2}$~s$^{-1}$; they are within a factor of few
agreement with our model calculations.  The weaker 15.55\,$\mu$m of
Ne\,III line has been tentatively detected in only one case (Sz\,102,
Lahuis et al.~2007). Based on the 3$\sigma$ detection of the Ne\,III
line, the Ne\,III/Ne\,II ratio is $\approx 0.06$, which might be
compared with a ratio $\sim 0.1$ derived from Table 3.  Perhaps what
is most interesting about this T Tauri star is that it has the second
brightest Ne\,II line detected in the Lahuis et al.~survey. Its flux
is an order of magnitude larger than the value for our reference
model, after correction for the distance of 200\,pc.  Detailed
modeling of the disk around this star would be of considerable
interest. For the case of nearby TW Hya, \citet{Herczeg2007} measure
the luminosity for the Ne\,II 12.81\,$\mu$m line to be $4.8 \times
10^{-6}\,L_{\odot}$, which is 20\% larger than the result for our
reference model in Table 3.

From a small number of detections of the Ne\,II 12.81 $\mu$m line the
{\it Spitzer} Lagacy program, ``Formation and Evolution of Planetary
Systems'' (FEPS), \citet{Pascucci2007} have suggested that a linear
correlation may exist between the Ne\,II and X-ray
luminosities. Together with their detection of the Ne\,II line in CS
Cha and that for TW Hya, based on data obtained by \citet{Uchida2004},
\citet{Espaillat2007} argue against such a correlation.  The range of
$L_X$ in these papers is small, 1.5 for \citet{Pascucci2007} and 2.25
for \citet{Espaillat2007}, the same order of magnitude as the observed
variability in the X-ray luminosity.  Aside from the intrinsic
difficulty of deducing a meaningful correlation on the basis of a
small dynamical range in YSO X-ray luminosity, it is important to keep
in mind that the line fluxes depend on many other stellar and disk
properties, e.g., the stellar temperature, mass, and age and the disk
mass, accretion rate, and the amount of dust grain growth and
settling. The theoretical calculations shown in Fig.~\ref{xray_trends}
may not be relevant in this context because they assume that the disk
structure is fixed while $L_X$ is varied. To pursue the question of
the existence of a correlation with $L_X$, T Tauri stars with similar
properties, both stellar and disk, should be considered.  A more
practical approach might be to focus in depth on a few specific cases
with well observed properties and to use observed X-ray spectra.

Using the MICHELLE spectrometer at Gemini North, \citet{Herczeg2007}
have obtained a spectrally resolved Ne\,II line profile for TW Hya,
which has a nearly face-on disk. The line is centered at the stellar
radial velocity, consistent with a disk origin, but the line is broad
with FWHM of $\sim 22$\,km s$^{-1}$. The rotational broadening in our
models (Fig.~5 of GNI07 and Fig.~\ref{neon_distribution} of this
paper), when corrected for inclination, cannot explain the observed
line width, even though the present calculations extend to radii as
small as 0.25\,AU. Although our calculations agree with the measured
Ne\,II luminosity, the rotational broadening in the model is small
because it arises from relatively large radii, as shown in
Fig.~\ref{neon_distribution}. As discussed by Herczeg et al. (2007),
two interesting explanations of the line width seen TW Hya that
deserve consideration are transonic turbulence and photo-evaporative
outflows from the disk \citep{Hollenbach2000,Font2004}. These
possibilities raise interesting challenges for future research. If the
hole is produced by photo-evaporation, then the wind itself will
generate line emission. Furthermore, the inner rim of the hole will be
irradiated more or less directly by the star, as in the model of
\citet{Chiang2007}, again producing characteristic line emission.

\subsubsection{Sulfur Fine-Structure Lines}

We consider the S\,I fine-structure lines to be ``high-excitation''
because the associated excitation energies, 476\,K and 825\,K, are
larger than the temperatures characteristic of the outer disk, as are
the fine-structure levels of the neon ions. As discussed in Section 3,
the X-ray ionization theory for neon and sulfur, where charge exchange
plays an important role, are also similar. Reference to Table 3 shows
that the integrated S\,I fine-structure emission for our reference
model is $\sim 20$ smaller than for the Ne\,II
line. Fig.~\ref{SI_finestruct} shows that the strongest emission of
the S\,I 25.25\,$\mu$m line comes from a relatively thin layer between
$N_{\rm H} = 10^{19}-10^{21}$\,cm$^{-2}$ for $R<10$\,AU. This
situation and the attendant weakness of the 25.25\,$\mu$m line occur
because sulfur is in the form of S$^+$ at higher altitudes and
electronic excitation becomes weak at lower altitudes
(c.f.~Figs.~\ref{sulfur_molecules} and \ref{S_fraction}).

The S\,I 25.25\,$\mu$m line was not detected by the {\it Spitzer} c2d
and FEPS teams.  \citet{Pascucci2007} put upper limits on the residual
gas in the disks around their survey of older T Tauri stars with ages
between 5 and 100\,Myr using the model calculations of
\citet{Gorti2004}. From Table 3, which is based on pure electronic
excitation, the S\,I fine-structure lines from younger T Tauri disks
would be similarly difficult to detect. However, if the atomic H
collision rates for S\,I are similar to those for O\,I, the predicted
fluxes would approach the observable range. It is important to recall
several other caveats that pertain to our calculations: the
uncertainties in sulfur chemistry, the total abundance of gaseous
sulfur in disks, and the poorly known charge-exchange rate
coefficients.

\subsubsection{Forbidden Lines}

Table \ref{integrated_luminosities} gives the fluxes for our reference
model of a sample of forbidden lines, as calculated in Section 4: O\,I
6300/6363\,\AA, O\,I 5577\,\AA, S\,II 6718/6731\,\AA; C\,I
9827/9853\,\AA. We have identified them as potential diagnostics of
the warm and highly-ionized gas produced by X-ray irradiation in the
upper atmosphere of the inner disk. These lines are generated within
the same radial distance range as the Ne\,II 12.81 and 15.55\,$\mu$m
lines, but at somewhat higher elevations. In principle, they can
provide complementary checks on our picture of X-ray irradiation. As
discussed in Section 4, the accuracy of the calculations varies
somewhat from line to line, depending on the reliability of the atomic
data and on the completeness of the model itself. For example, we have
ignored the role of stellar or interstellar/intracluster FUV
radiation, which could enhance the ionization of carbon at high
altitudes and thereby reduce the amount of neutral carbon and the
strength of the C\,I forbidden lines. Similarly, the S\,II 6718\,\AA\
line strength is affected by the sulfur atoms being converted into
molecules.  With this proviso, we can see from
Table~\ref{integrated_luminosities} that many of the forbidden lines
generated in the upper disk atmosphere are observable in nearby
star-forming regions such as Taurus-Aurigae in the sense that the
predicted fluxes are $\gtrsim 10^{-15}$erg\, cm$^{-2}$\, s$^{-1}$,
notably O\,I 6300/6363\,\AA, C\,I 9827/9853\,\AA, and perhaps S\,I
6718/6733\,\AA.

Many of these lines have been measured in T Tauri stars. The most
dramatic manifestation of the forbidden lines is their tracing of
high-velocity outflows or jets from YSOs (e.g., Ray et
al.~\citeyear{Ray2007}). But these flows have a low-velocity as well
as a high-velocity component
\citep{Kwan1988,Kwan1995,Hirth1997}. \citet{Kwan1997} suggested that
the low velocity component (LVC) originates in the inner regions of T
Tauri disks ($R < 2$\,AU)\footnote{Kwan's theory of warm disk coronae
has similarities with our X-ray irradiated disk atmospheres, but there
are significant differences in the underlying ionization physics.}.
Other possible sources for the origin of the LVC is a MHD disk wind
(e.g., Pudritz et al.~\citeyear{Pudritz2007}) or a photoevaporative
outflow (e.g., Hollenbach et al.~\citeyear{Hollenbach2000}; Font et
al.~\citeyear{Font2004}).

\citet{Hartigan1995} have obtained extensive data on forbidden line
emission from T Tauri stars. The LVC is ubiquitous, with a typical
line luminosity of $\sim 10^{-4}\,L_{\odot}$, an order of magnitude
larger than the luminosity of our reference model, $8.0 \times
10^{-6}\,L_{\odot}$. They also find O\,I 5577\,\AA\ to O\,I 6300\,\AA\
line ratios in the range $\sim 0.2-0.5$; our model predicts $\sim
0.1$. These numbers may signify the difficulty of applying our X-ray
irradiated disk model to observations such as the optical forbidden
lines that are sensitive to the other dynamic entities that are
operative in accreting young stars, e.g., the various flows that arise
near the inner edge of the disk. Hartigan et al.~define the LVC to
include velocities as high as 60 km s$^{-1}$, thereby increasing the
possibility of including emission from such flows. A cleaner
comparison case for our model might be provided by an X-ray bright
star with a relatively low accretion rate, where the emission from
outflows would be reduced.

Because TW Hya has a low mass-loss rate and a small accretion rate, it
would be a good case for detailed modeling. \citet{Herczeg2007}
detected the O\,I 6300/6363\,\AA\,lines in TW Hya at moderately high
spectral resolution. The luminosity of the 6300\,\AA\,line, $7.0
\times 10^{-6}\,L_{\odot}$, is close to the result given in Table 3,
$8.0 \times 10^{-6}\,L_{\odot}$, recalling the good agreement for the
luminosity of the Ne\,II 12.81\,$\mu$m line. The 6300\,\AA\,line is
centered on the stellar velocity, and it has a significantly narrower
width than the Ne\,II line, but one that is still too large to be
explained by pure rotational broadening. Again, interesting possible
explanations to consider in future modeling are a turbulent atmosphere
and photo-evaporation.

\subsection{Low excitation lines}

We next consider the low-excitation lines that arise from the
fine-structure levels of O\,I (excitation energies 228\,K and 327\,K),
C\,I (excitation energies 23.6\,K and 62.5\,K) and C\,II (excitation
energy 91.2\,K); they give rise to far-infrared emission at 63 and
145\,$\mu$m (O\,I), 609 and 370\,$\mu$m (C\,I), and 158\,$\mu$m
(C\,II). Table 3 shows that the O\,I fine-structure lines are
potential diagnostics of both the moderately warm, ionized gas
produced by X-rays, and the cooler gas at large perpendicular column
densities. As shown in Fig.~\ref{OI_finestruct}, O\,I 63\,$\mu$m line
emission extends down toward the mid-plane region, especially at small
radial distances, due to the incomplete conversion of atomic oxygen
into molecules and the relatively low excitation temperature of the
$J=1-2$ transition.

The 63\,$\mu$m line has been observed around T Tauri stars with the
KAO (e.g., Cohen et al.~\citeyear{Cohen1988}; Ceccarelli et
al.~\citeyear{Ceccarelli1997}) and ISO (e.g, Spinogolio et
al.~\citeyear{Spinoglio2000}; Creech-Eakman et
al.~\citeyear{Creech2002}; Liseau et al. \citeyear{Liseau2006}). These
observations were made with large beams (tens of arc seconds) and with
moderate spatial resolution (50-300 km s$^{-1}$). A typical flux for a
T Tauri star in Taurus-Aurigae is quite large, $\sim
10^{-12}-10^{-11}$\,erg cm$^{-2}$\, s$^{-1}$, and may include emission
from other circumstellar material and not just disks. On the other
hand, the flux calculations in Table 3 probably underestimate the
flux, even though they integrate out to 100\,AU, especially for the
carbon lines. \citet{Jonkheid2004} have applied a code developed for
(UV) photon-dominated regions to the surface layers of a flaring disk,
and find that the fluxes of the O\,I, C\,I, and C\,II fine-structure
lines are all of comparable brightness, and not that much weaker than
the CO($J=1-0$) line. Photo-ionization of neutral carbon by external
FUV radiation can enhance the abundance of C$^+$ and increase the flux
of the C\,II 158\,$\mu$m line. Our model would have to be extended to
include UV irradiation in order to get a fuller understanding of these
low-excitation diagnostic lines.

Our estimate of the O\,I 63\,$\mu$m line flux provides a good estimate
of the emission of this line from the inner disk. Since its detection
requires airborne or space observations, the identification with inner
disk emission requires resolved line profiles and the assumption of
Keplerian disk rotation.  Such measurements could be made with
heterodyne detectors aboard SOFIA or with {\it Herschel} (PACS).
 
\section{Summary}

We have used a simplified thermal-chemical and structural model to
explore the diagnostics of the atomic gas in protoplanetary disks
irradiated by stellar X-rays. Our focus has been on moderate disk
radii out to roughly 25\,AU, where stellar X-rays play an important
role in ionizing and heating the upper layers of the disk.  We find
several diagnostics for this region, all of which arise from
relatively high-excitation levels $\sim 1000$\,K: Ne\,II and Ne\,III
fine-structure lines and O\,I, S\,II, and C\,I forbidden lines.  Some
low-excitation fine-structure lines are also of interest in the
context of X-ray irradiation, especially the O\,I fine-structure
lines, which generates strong disk emission over a substantial range
of vertical column density, extending close to the mid-plane of the
disk.

The atomic lines probe different depths of the disk at moderate
radii. The forbidden transitions are generated at the top of the
atmosphere, where the temperature and ionization levels are the
highest. The mid-infrared fine-structure lines probe greater depths,
where the temperature and ionization level have decreased. Neon is the
cleanest case, since it does not easily form molecules, and its
fine-structure lines originate from intermediate as well as top layers
of the disk. On the other hand, the S\,I fine-structure lines are
produced mainly in intermediate layers, since atomic sulfur has a low
abundance in the top layers due to X-ray ionization and also near the
mid plane, due to molecule formation. By contrast, the O\,I
fine-structure lines are formed at almost all depths, including close
to the mid plane, where the density is high and the upper levels of
the transitions are more commensurate with the temperatures there.

The X-ray generated atomic lines discussed here are produced over a
range of radii, extending from the inner radius of the disk out to
20-30\,AU. To resolve these regions in nearby star-forming clusters at
a nominal distance of 140\,pc, requires a spatial resolution of $\sim
0.1-0.2$\arcsec, as well as resolved line profiles. For example, the
Ne\,II 12.81\, $\mu$m line should be detectable with TEXES or
comparable spectrometers on large ground-based telescopes. A promising
step in this direction is its detection with MICHELLE on Gemini North
\citep{Herczeg2007}.  The far-infrared lines of C\,I, C\,II, and O\,I
will be detectable with instruments on board SOFIA and {\it Herschel}.

In presenting these results for a simplified T Tauri disk model
irradiated by stellar X-rays, we have tried to mention and discuss its
limitations. To summarize, there are, first of all, many uncertainties
in the underlying atomic physics, such as missing rate coefficients
for charge exchange with atomic hydrogen and for collisional
excitation by abundant particles other than electrons. These
uncertainties of course hold for all thermal-chemical/excitation
models. Another limitation is the reliance on one type of disk,
represented by the smooth density distribution of the generic T Tauri
model of D'Alessio et al.~(1999). Removing this restriction is one of
our immediate goals. It will enable us to discuss disks of different
ages and disks with non-monotonic density distributions such as holes,
gaps, and rims. The improvements that are needed to treat situations
such as disks with holes require a higher level model than used
here. More generally, a 3-dimensional treatment is needed, and this
dictates a greatly improved treatment of the radiation transfer,
especially for the cooling lines and the external UV irradiation.

Focusing on X-rays to the exclusion of UV irradiation is another
limitation that needs to be dealt with in future modeling. It is very
likely that the well-documented X-rays produced by essentially all
YSOs are the dominant external radiation source for radii less than
25\,AU. At larger radii, however, ambient UV, from the general
interstellar medium or from the host star cluster members, can affect
disk properties. At smaller radii, UV radiation originating close to
the star may also play a role, although it is more easily attenuated
than moderately hard keV X-rays. We have ignored stellar UV radiation,
which is poorly determined, in favor of the well-measured X-rays.

Granted that the X-rays likely play the dominant role within 25\,AU or
so, the relatively high-excitation transitions that they generate may
be produced by some of the other dynamical components of the
star-formation system that lie near and even inside the inner edge of
the disk. The most obvious examples are the forbidden optical
transitions produced by jets and accretion streams that are the
signatures of actively accreting YSOs. In this situation, the
high-excitation line strengths that we calculate for disks may only
represent lower limits to those measured in spatially unresolved
observations. A similar conclusion pertains to the low-excitation
lines, where significant emission can arise from large disk radii
where external UV can dominate. Of course the contributions of the
various parts of the star formation complex can in principle be
disentangled by observations with the appropriate spatial and spectral
resolution, together with more complex models. The present results
demonstrate the importance of X-ray irradiation in the ongoing process
of elucidating the nature of the gas in protoplanetary disks.

\section{Acknowledgements} 

This work has been supported by NSF Grant AST-0507423 and NASA Grant
NNG06GF88G. We would like to express our appreciation to Dr.~Javier 
Igea for his development of the thermal-chemical program used in this 
research. We would also like to thank Dr.~Paola D'Alessio for providing 
the results of her disk model which form the basis of the calculations 
reported in this and previous publications.

\clearpage

\clearpage

\rotate
\pagestyle{empty}
\begin{table*}
\caption{Selected Reaction Rate coefficients for Sulfur}
\centering
\begin{tabular}{|l|l|l|l|l|}
\tableline
Reactants & Products & Symbol & Rate coefficient & Reference \\
\tableline
${\rm S}^{+} + {\rm e}$  & ${\rm S} + h\nu$       & $\alpha_1$ & $4.65\times 10^{-13} T_4^{-0.63}$ & \citet{Aldrovandi1973, Gould1978} \\
${\rm S}^{2+} + {\rm e}$ & ${\rm S}^{+} + h\nu$   & $\alpha_2$ & $3.37\times 10^{-12} T_4^{-0.50}$ & \citet{Nahar1995} \\
${\rm S}^{3+} + {\rm e}$ &  ${\rm S}^{2+} + h\nu$ & $\alpha_3$ & $5.23\times 10^{-12} T_4^{-0.50}$ & \citet{Nahar1995} \\
\tableline
${\rm H}^{+} + {\rm S}$  & ${\rm S}^{+} + {\rm H}$      & $k_1$ & $5\times 10^{-12}$* & \citet{Zhao2005} \\
${\rm S}^{+} + {\rm H}$  & ${\rm H}^{+} + {\rm S}$      & $k_1$ & negligible          & \citet{Zhao2005} \\
${\rm S}^{2+} + {\rm H}$ & ${\rm H}^{+} + {\rm S}^+$    & $k_2$ & $10^{-14}$          & \citet{Butler1980, Christensen1981} \\
${\rm S}^{3+} + {\rm H}$ & ${\rm H}^{+} + {\rm S}^{2+}$ & $k_3$ & $3\times 10^{-9}$   & \citet{Butler1980, Christensen1981} \\
\tableline
${\rm S}^{2+} + {\rm H_2}$ & ${\rm S}$ etc. & $BK$      & $2\times 10^{-9}$B         & \citet{Chen2003} \\
${\rm S}^{2+} + {\rm H_2}$ & ${\rm S}^{+}$ etc. & $(1-B)K$ & $2\times 10^{-9}$(1-B)     & \citet{Chen2003} \\
\tableline
${\rm S} + $ X-ray & ${\rm S^+} + e$ & $\zeta_{\rm sec}({\rm S})$ &  $5\times \zeta$ & 
This work.\\
${\rm S} + $ X-ray & ${\rm S^{2+}} + 2e$ & $\zeta_{\rm dir}({\rm S})$ &  $20\times \zeta$ & 
This work.\\
${\rm S}^+ + $ X-ray & ${\rm S^{2+}} + e$ & $\zeta({\rm S})$ &  $25\times \zeta$ &  
This work.\\
\tableline
\multicolumn{5}{l}{* Approximate value valid from 500-5,000\,K}\\

\end{tabular}
\label{rates}

\end{table*}

\clearpage

\begin{table*}
\caption[]{References for excitation coefficients*}
\centering
\begin{tabular}{|l|l|}
\tableline
Ion & references   \\        
\tableline
Ne\,II fine structure   & \citet{Griffin2001} (e) \\
Ne\,III fine structure  & \citet{Butler1994} (e) \\       
O\,I fine structure     & \citet{Launay1977a}; \citet{Abrahamsson2007} (H); 
			  \citet{Jaquet1992} (H$_2$); \\
                        & \citet{Monteiro1987} (He); \citet{Chambaud1980} (p); 
			   \citet{Pequignot1990} (e) \\
O\,I forbidden      & \citet{Krems2006} (H); \citet{Zatsarinny2003} (e) \\  
S\,I fine structure             & \citet{Tayal2004} (e)      \\
S\,II forbidden     & \citet{Tayal1997} (e)      \\
C\,I fine structure     & \citet{Launay1977a}; \citet{Abrahamsson2007} (H); \\
                  & \citet{Monteiro1987} (He, H$_2$) \\
C\,I forbidden      & \citet{Pequignot1976}; \citet{Zatsarinny2005} (e) \\
C\,II fine structure  & \citet{Barinovs2005} (H); \citet{Launay1977b} (H$_2$); 
		       \citet{Wilson2002} (e) \\
\tableline
\multicolumn{2}{l}{*The collision partners are specified inside the brackets.}\\
\end{tabular}
\label{excitation_refs}
\end{table*}

\clearpage

\begin{table*}
\caption[]{Unresolved line fluxes in erg\,s$^{-1}$\,cm$^{-2}$*}
\centering
\begin{tabular}{|l|l|l|l|l|l|}
\tableline
$L_X$ (erg~s$^{-1}$) & $2.0\times 10^{29}$ &  $6.0\times 10^{29}$ & $\mathbf{2.0\times 10^{30}}$ & $6.0\times 10^{30}$ &  $2.0\times 10^{31}$ \\
\tableline
Ne\,II 12.82\,$\mu$m  & $4.1\times 10^{-16}$ & $1.5\times 10^{-15}$ & $\mathbf{6.4\times 10^{-15}}$ & $2.3\times 10^{-14}$ & $9.6\times 10^{-14}$ \\ 
Ne\,III 15.55\,$\mu$m & $3.0\times 10^{-17}$ & $1.4\times 10^{-16}$ & $\mathbf{7.4\times 10^{-16}}$ & $3.3\times 10^{-15}$ & $1.7\times 10^{-14}$ \\
\tableline
O\,I 63\,$\mu$m       & $1.9\times 10^{-14}$ & $3.1\times 10^{-14}$ & $\mathbf{6.6\times 10^{-14}}$ & $1.6\times 10^{-13}$ & $4.3\times 10^{-13}$ \\
O\,I 146\,$\mu$m      & $9.6\times 10^{-16}$ & $1.2\times 10^{-15}$ & $\mathbf{1.8\times 10^{-15}}$ & $3.8\times 10^{-15}$ & $1.1\times 10^{-14}$ \\
O\,I 6300\,\AA **     & $1.1\times 10^{-15}$ & $3.6\times 10^{-15}$ & $\mathbf{1.3\times 10^{-14}}$ & $4.5\times 10^{-14}$ & $1.7\times 10^{-13}$ \\
O\,I 5577\,\AA        & $2.7\times 10^{-17}$ & $1.4\times 10^{-16}$ & $\mathbf{7.3\times 10^{-16}}$ & $2.8\times 10^{-15}$ & $1.1\times 10^{-14}$ \\
\tableline
S\,I 25.25\,$\mu$m    & $7.0\times 10^{-17}$ & $1.3\times 10^{-16}$ & $\mathbf{3.4\times 10^{-16}}$ & $9.5\times 10^{-16}$ & $3.1\times 10^{-15}$ \\
S\,I 56.23\,$\mu$m    & $6.4\times 10^{-18}$ & $9.0\times 10^{-18}$ & $\mathbf{1.7\times 10^{-17}}$ & $4.0\times 10^{-17}$ & $1.2\times 10^{-16}$ \\
S\,II 6718\,\AA       & $4.1\times 10^{-17}$ & $1.7\times 10^{-16}$ & $\mathbf{8.3\times 10^{-16}}$ & $3.5\times 10^{-15}$ & $1.6\times 10^{-14}$ \\
S\,II 6733\,\AA       & $9.2\times 10^{-18}$ & $3.9\times 10^{-17}$ & $\mathbf{2.0\times 10^{-16}}$ & $8.6\times 10^{-16}$ & $4.2\times 10^{-15}$ \\
\tableline
C\,II 158\,$\mu$m     & $1.2\times 10^{-18}$ & $3.2\times 10^{-18}$ & $\mathbf{9.8\times 10^{-18}}$ & $2.8\times 10^{-17}$ & $9.0\times 10^{-17}$ \\
C\,I 369\,$\mu$m      & $9.8\times 10^{-16}$ & $1.2\times 10^{-15}$ & $\mathbf{1.6\times 10^{-15}}$ & $2.0\times 10^{-15}$ & $2.6\times 10^{-15}$ \\
C\,I 609\,$\mu$m      & $3.2\times 10^{-16}$ & $3.8\times 10^{-16}$ & $\mathbf{4.5\times 10^{-16}}$ & $5.4\times 10^{-16}$ & $6.4\times 10^{-16}$ \\
C\,I 9827\,\AA **     & $3.5\times 10^{-16}$ & $1.4\times 10^{-15}$ & $\mathbf{6.3\times 10^{-15}}$ & $2.5\times 10^{-14}$ & $1.1\times 10^{-13}$ \\
\tableline
\multicolumn{6}{l}{* A nominal distance of 140 pc has been assumed. 
Luminosities can be obtained by}\\
\multicolumn{6}{l}{multiplying by $4\pi (140 {\rm pc})^2 = 
2.35 \times 10^{42}$\,cm$^2$. The middle column refers to the}\\
\multicolumn{6}{l}{reference model.}\\   
\multicolumn{6}{l}{** The O\,I 6363\,\AA\ and C\,I 9853\,\AA\ fluxes 
can be obtained by multiplying the fluxes of} \\
\multicolumn{6}{l}{O\,I 6300\,\AA\ by 0.244 and C\,I 9827\,\AA\ by 0.248, respectively.}\\  
\end{tabular}
\label{integrated_luminosities}
\end{table*}

\end{document}